\documentclass[aps,eqsecnum,preprint,floats,epsf,epsfig,nofootinbib]{revtex4}
\usepackage{epsfig}
\usepackage{graphicx}% Include figure files

\begin{document}
\def\be{\begin{eqnarray}}
\def\en{\end{eqnarray}}
\def\non{\nonumber}
\def\la{\langle}
\def\ra{\rangle}
\def\A{{\cal A}}
\def\B{{\cal B}}
\def\c{{\cal C}}
\def\d{{\cal D}}
\def\e{{\cal E}}
\def\p{{\cal P}}
\def\t{{\cal T}}
\def\nc{N_c^{\rm eff}}
\def\CP{{\it CP}~}
\def\CPP{{\it CP}}
\def\vp{\varepsilon}
\def\drho{\bar\rho}
\def\deta{\bar\eta}
\def\vma{{_{V-A}}}
\def\vpa{{_{V+A}}}
\def\J{{J/\psi}}
\def\ov{\overline}
\def\Lqcd{{\Lambda_{\rm QCD}}}
\def\pr{{ Phys. Rev.}~}
\def\prl{{ Phys. Rev. Lett.}~}
\def\pl{{ Phys. Lett.}~}
\def\np{{ Nucl. Phys.}~}
\def\zp{{ Z. Phys.}~}
\def\lsim{ {\ \lower-1.2pt\vbox{\hbox{\rlap{$<$}\lower5pt\vbox{\hbox{$\sim$}
}}}\ } }
\def\gsim{ {\ \lower-1.2pt\vbox{\hbox{\rlap{$>$}\lower5pt\vbox{\hbox{$\sim$}
}}}\ } }

\font\el=cmbx10 scaled \magstep2{\obeylines \hfill July, 2016}

\vskip 1.0 cm

\centerline{\large\bf Direct {\it CP} Violation}
\centerline{\large\bf in Charmless Three-body Decays of $B$ Mesons}
\bigskip
\centerline{\bf Hai-Yang Cheng$^{1}$, Chun-Khiang Chua$^{2}$, Zhi-Qing Zhang$^{3}$}
\medskip
\centerline{$^1$ Institute of Physics, Academia Sinica}
\centerline{Taipei, Taiwan 115, Republic of China}
\medskip
\centerline{$^2$ Department of Physics and Center for High Energy Physics}
\centerline{Chung Yuan Christian University}
\centerline{Chung-Li, Taiwan 320, Republic of China}
\medskip
\centerline{$^3$ Department of Physics, Henan University of Technology} \centerline{Zhengzhou, Henan 450052, P.R.
China}
\bigskip
%\medskip

\bigskip
\bigskip
\centerline{\bf Abstract}
\bigskip

\small

Direct \CP violation in
charmless three-body hadronic decays of $B$ mesons is studied within the framework of a simple model based on the factorization approach.  Three-body decays of heavy mesons receive both resonant and nonresonant contributions.
Dominant nonresonant contributions to tree-dominated and penguin-dominated three-body decays arise from the $b\to u$ tree transition and $b\to s$ penguin transition, respectively. The former can be evaluated in the framework of  heavy meson chiral perturbation theory with some modification, while the latter is
governed by the matrix element of the scalar density $\la M_1M_2|\bar q_1 q_2|0\ra$.
Resonant contributions to three-body decays are treated using the isobar model. Strong phases in this work reside in effective Wilson
coefficients, propagators of resonances and the matrix element of scalar density.
In order to accommodate the branching fraction and \CP asymmetries observed in $B^-\to K^-\pi^+\pi^-$, the matrix element $\la K\pi|\bar sq|0\ra$ should have an additional strong phase, which might arise from some sort of power corrections such as final-state interactions. We calculate inclusive and regional \CP asymmetries and find that nonresonant \CP violation is usually much larger than the resonant one and that the interference effect between resonant and nonresonant components is generally quite significant.
If nonresonant contributions are turned off in the $K^+K^-K^-$ mode, the predicted \CP asymmetries due to resonances will be wrong in sign when confronted with experiment.
In our study of $B^-\to \pi^-\pi^+\pi^-$, we find that $\A_{C\!P}(\rho^0\pi^-)$ should be positive in order to account for \CP asymmetries observed in this decay. Indeed, both BaBar and LHCb measurements of $B^-\to \pi^+\pi^-\pi^-$ indicate positive \CP asymmetry in the $m(\pi^+\pi^-)$ region peaked at $m_\rho$. On the other hand, all theories predict a large and negative \CP violation in $B^-\to \rho^0\pi^-$. Therefore, the issue with \CP violation in $B^-\to\rho^0\pi^-$ needs to be resolved. Measurements of {\it CP}-asymmetry Dalitz distributions put very stringent constraints on the theoretical models. We check the magnitude and the sign of \CP violation in some (large) invariant mass regions to test our model.

\pagebreak

%%%%%%%%%%%%%%%%%%%%%
\section{Introduction}
%%%%%%%%%%%%%%%%%%%%%

The primary goal and the most important mission of $B$ factories built before millennium is to search for \CP violation in the $B$ meson system. BaBar and Belle have measured direct \CP asymmetries in many two-body charmless hadornic $B$ decay channels, but only ten of them have significance large than 3$\sigma$: $ B^-/\ov B^0\to  K^-\pi^+, \pi^+\pi^-, K^-\eta, \ov K^{*0}\eta, K^{*-}\pi^+, K^-f_2(1270), \pi^-f_0(1370), K^-\rho^0, \rho^\pm\pi^\mp$ \cite{PDG,HFAG} and $B^-\to K^{*-}\pi^0$ \cite{BaBar:Kstpi}.
In the $B_s$ system, direct \CP violation in $\bar B_s^0\to K^+\pi^-$ with $7.2\sigma$ significance was measured by LHCb \cite{LHCb:Bs}.
As for three-body $B$ decays, BaBar and Belle had measured partial rate asymmetries in various charmless three-body modes (see \cite{PDG,HFAG} or Table I of \cite{Cheng:2013dua}), and failed no see any evidence.

Recently, LHCb has measured direct \CP violation in charmless three-body decays of $B$ mesons
\cite{LHCb:Kppippim,LHCb:pippippim,LHCb:2014} and found evidence of inclusive integrated \CP asymmetries $\A_{C\!P}^{\rm incl}$ in $B^+\to\pi^+\pi^+\pi^-$
(4.2$\sigma$), $B^+\to K^+K^+K^-$ (4.3$\sigma$) and $B^+\to K^+K^-\pi^+$ (5.6$\sigma$) and a 2.8$\sigma$ signal of
\CP violation in $B^+\to K^+\pi^+\pi^-$ (see Table \ref{tab:CPdata}). Direct \CP violation in two-body resonances in the
Dalitz plot has been seen at $B$ factories. For example, both BaBar \cite{BaBar:Kmpippim} and Belle
\cite{Belle:Kmpippim} have claimed evidence of partial rate asymmetries in the channel $B^\pm\to \rho^0(770)K^\pm$ in
the Dalitz-plot analysis of $B^\pm\to K^\pm\pi^\mp\pi^\pm$.
The inclusive \CP asymmetry in three-body decays results from the interference of the two-body resonances and three-body
nonresonant decays and from the tree-penguin interference. \CP asymmetries in certain local regions of the phase space
are likely to be greater than the integrated ones. Indeed, LHCb has also observed large asymmetries in localized
regions of phase space  (see Table \ref{tab:CPdata} for $\A_{C\!P}^{\rm
low}$) specified by \cite{LHCb:Kppippim,LHCb:pippippim}
\be \label{eq:KKKlocalCP}
\A_{C\!P}^{\rm low}(K^+K^-K^-),&&~{\rm for}~m^2_{K^+K^- \rm ~high}<15~{\rm GeV}^2, ~1.2<m^2_{K^+K^- \rm
~low}<2.0~{\rm GeV}^2, \non \\
\A_{C\!P}^{\rm low}(K^-\pi^+\pi^-),&& ~
{\rm for}~ m^2_{K^-\pi^+ \rm ~high}<15~{\rm GeV}^2,~ 0.08<m^2_{\pi^+\pi^- \rm ~low}<0.66~{\rm GeV}^2, \non \\
\A_{C\!P}^{\rm low}(K^+K^-\pi^-), && ~{\rm for}~ m^2_{K^+K^-}<1.5~{\rm GeV}^2, \\
\A_{C\!P}^{\rm low}(\pi^+\pi^-\pi^-),&&~
{\rm for}~ m^2_{\pi^+\pi^- \rm ~low}<0.4~{\rm GeV}^2,~m^2_{\pi^+\pi^- \rm ~high}>15~{\rm GeV}^2. \non
\en
Hence, significant signatures of \CP violation were found in the above-mentioned low mass regions devoid of most of the
known resonances.
LHCb has also studied \CP asymmetries in the rescattering regions of $m_{\pi^+\pi^-}$ or $m_{K^+K^-}$ between 1.0 and 1.5 GeV where the final-state $\pi^+\pi^-\leftrightarrow K^+K^-$ rescattering is supposed to be important in this region. The measured \CP asymmetries $\A_{C\!P}^{\rm
resc}$ for the charged final states are given in Table \ref{tab:CPdata}.

In two-body $B$ decays, the measured \CP violation is just a number. But in three-body decays, one can measure the distribution of \CP asymmetry in the Dalitz plot. Hence, the Dalitz-plot analysis of $\A_{C\!P}$ distributions can reveal very rich information about \CP violation. Besides the integrated \CP asymmetry, local asymmetry can be very large and positive in some region and becomes very negative in the other region. The sign of \CP asymmetries varies from region to region. A successful model must explain not only the inclusive asymmetry but also regional \CP violation. Therefore, the study of three-body {\it CP}-asymmetry Dalitz distributions provides a great challenge to the theorists.
LHCb has measured the raw asymmetry $A_{\rm raw}$ distributions in the Dalitz plots  defined by \cite{LHCb:2014}
\be
A_{\rm raw}={N_{B^-}-N_{B^+}\over   N_{B^-}+N_{B^+}}
\en
in terms of numbers of $B^-$ and $B^+$ signal events $N_{B^-}$ and $N_{B^+}$, respectively. The relation between $A_{\rm raw}$ and $\A_{C\!P}$ is given in
\cite{LHCb:Kppippim,LHCb:pippippim,LHCb:2014}.  Two-body invariant-mass projection plots are available in Figs. 4--7 of \cite{LHCb:2014}. For \CP Dalitz asymmetries in high invariant mass regions, see \cite{LHCb:2016}.

Three-body decays of heavy mesons are more complicated than the
two-body case as they receive both resonant and nonresonant
contributions.  The
analysis of these decays using the Dalitz plot technique enables
one to study the properties of various vector and scalar resonances.
Indeed, most of the quasi-two-body decays are
extracted from the Dalitz-plot analysis of three-body ones.  In this work we shall focus on  charmless $B$
decays into three pseudoscalar mesons.

%%%%%%%%%%%%%%%%%%%%%%%%%%%%%%%%%%%
\begin{table}[t]
\caption{LHCb results of direct \CP asymmetries (in \%) for various charmless
three-body $B^-$ decays. The superscripts ``incl", ``low" and ``resc" denote \CP asymmetries measured in full phase space, in the low invariant mass regions specified in Eq. (\ref{eq:KKKlocalCP}) and in the rescattering regions with 1.0 $<m_{\pi^+\pi^-,K^+K^-}<$ 1.5 GeV, respectively. Data are taken from  \cite{LHCb:Kppippim,LHCb:pippippim} for $\A_{C\!P}^{\rm low}$  and from \cite{LHCb:2014} for $\A_{C\!P}^{\rm incl}$ and $\A_{C\!P}^{\rm resc}$. }
\begin{ruledtabular} \label{tab:CPdata}
\begin{tabular}{l  c c c c }
  & $\pi^+\pi^-\pi^-$
  & $K^+K^-\pi^-$
  & $K^-\pi^+\pi^-$
  & $K^-K^+K^-$
  \\ \hline
$\A_{C\!P}^{\rm incl}$
  & $5.8\pm0.8\pm0.9\pm0.7$
  & $-12.3\pm1.7\pm1.2\pm0.7$
  & $2.5\pm0.4\pm0.4\pm0.7$
  & $-3.6\pm0.4\pm0.2\pm0.7$ \\
$\A_{C\!P}^{\rm low}$
  & $58.4\pm8.2\pm2.7\pm0.7$
  & $-64.8\pm7.0\pm1.3\pm0.7$
  & $67.8\pm7.8\pm3.2\pm0.7$
  & $-22.6\pm2.0\pm0.4\pm0.7$ \\
$\A_{C\!P}^{\rm resc}$
  & $17.2\pm2.1\pm1.5\pm0.7$
  & $-32.8\pm2.8\pm2.9\pm0.7$
  & $12.1\pm1.2\pm1.7\pm0.7$
  & $-21.1\pm1.1\pm0.4\pm0.7$ \\
\end{tabular}
\end{ruledtabular}
\end{table}
%%%%%%%%%%%%%%%%%%%%%%%%%%%%%%%%

Contrary to three-body $D$ decays where the nonresonant signal is usually rather small and less than 10\% \cite{PDG}, nonresonant contributions play an essential role in penguin-dominated three-body $B$ decays. For example, the nonresonant fraction of $KKK$ modes is of order (70-90)\%.
It follows  that nonresonant contributions to the penguin-dominated modes should be also dominated by the penguin mechanism.
It has been shown in \cite{CCS:nonres,Cheng:2013dua} that
large nonresonant signals arise mainly from the penguin amplitude governed by the matrix element of scalar densities $\la M_1M_2|\bar q_1 q_2|0\ra$.  We use the measurements of $\ov B^0\to K_SK_SK_S$  to constrain the nonresonant component of $\la K\ov K|\bar ss|0\ra$ \cite{CCS:nonres}.

Even for tree-dominated three-body decays such as $B^-\to \pi^-\pi^+\pi^-$, the nonresonant fraction is about 35\%.
In this case, dominant nonresonant contributions arise
from the $b\to u$ tree transition which can be evaluated using heavy meson chiral perturbation theory (HMChPT) \cite{Yan,Wise,Burdman} valid in the soft meson limit. The momentum dependence of nonresonant $b\to u$ transition amplitudes is parameterized in an exponential form
$e^{-\alpha_{_{\rm NR}} p_B\cdot(p_i+p_j)}$ so that the HMChPT
results are recovered in the soft meson limit where $p_i,~p_j\to 0$. The parameter $\alpha_{_{\rm NR}}$ is fixed by the measured nonresonant rate in $B^-\to\pi^+\pi^-\pi^-$.

Besides the nonresonant background, it is necessary to study resonant
contributions to three-body decays.
Resonant effects are conventionally described using the isobar model in terms of the usual Breit-Wigner formalism. In this manner we are able to
identify the relevant
resonances which contribute to the three-body decays of interest and
compute the rates of $B\to VP$ and $B\to SP$, where
the intermediate vector meson contributions to three-body decays are
identified through the vector current, while the scalar meson
resonances are mainly associated with the scalar density. They can also contribute to
the three-body matrix element $\la P_1P_2|J_\mu|B\ra$.

The recent LHCb measurements of integrated and local direct \CP asymmetries in charmless $B\to P_1P_2P_3$ decays (see Table \ref{tab:CPdata}) provide a new insight of the underlying mechanism of three-body decays. The observed negative relative sign of \CP asymmetries between $B^-\to \pi^-\pi^+\pi^-$
and $B^-\to K^-K^+K^-$ and between $B^-\to K^-\pi^+\pi^-$ and $B^-\to \pi^-K^+K^-$ is in accordance with what
expected from U-spin symmetry which enables us to relate the $\Delta S=0$ amplitude to the $\Delta S=1$ one. However,
symmetry arguments alone do not tell us the relative sign of \CP asymmetries between $\pi^-\pi^+\pi^-$ and $\pi^-K^+K^-
$ and between $K^-\pi^+\pi^-$ and  $K^-K^+K^-$.
The observed asymmetries (integrated or regional) by LHCb are positive for $h^-\pi^+\pi^-$ and negative for $h^-K^+K^-$ with $h=\pi$ or $K$. The former usually has a larger \CP asymmetry in magnitude than the latter. This
has led to the conjecture that $
\pi^+\pi^-\leftrightarrow K^+K^-$
rescattering may play an important role in the generation of the strong phase
difference needed for such a violation to occur \cite{LHCb:2014}.

After the LHCb measurement of direct \CP violation in  three-body charged $B$ decays, there are some theoretical
works in this regard \cite{Zhang,Bhattacharya,XGHe,Cheng:2013dua,Bediaga,Lesniak,Li:2014,Wang:2014ira,Bhattacharya:2014,Krankl:2015fha,Wang:2015ula,Bediaga:2015,Bediaga:2015mia}.
In the literature, almost all the works focus on resonant contributions to the rates and asymmetries. This is understandable in terms of the experimental observation that 90\% of the Dalitz plot events has $m(h^+h^-)^2<3.0\,{\rm GeV}^2$ \cite{Bediaga:Honnef}. The events are concentrated in low-mass regions, implying the dominance of charmless decays by resonant contributions.
Nevertheless,
in \cite{Cheng:2013dua} we have examined \CP violation in three-body decays and stressed the crucial role played by the nonresonant contributions. Indeed, if the nonresonant term is essential to account for the total rate, it should play some role to \CP violation. In this work, we would like to study asymmetries arising from both resonant and nonresonant amplitudes and their interference. This will make it clear the relative weight of both contributions and their interference.

It has been argued in \cite{Krankl:2015fha} that the amplitude at the Dalitz plot center is expected to be both power- and strong coupling $\alpha_s$-suppressed with respect to the amplitude at the edge. The perturbative regime in the central region gets considerably reduced for realistic value of $m_B$. That is, the Dalitz plot is completely dominated by the edges. Since the nonresonant background arises not just from the central region, the above argument is not inconsistent with the experimental observation of dominant nonresonant signals in penguin-dominated 3-body decays.

There are several competing approaches for describing charmless hadronic two-body decays of $B$ mesons, such as  QCD factorization
(QCDF) \cite{BBNS}, perturbative QCD (pQCD) \cite{Li} and soft-collinear
effective theory (SCET) \cite{SCET}. Unlike the two-body case, to date we still do not have theories for hadronic three-body decays, though attempts along the framework of pQCD and QCDF have been made in the past \cite{Chen:2002th,Wang:2014ira,Krankl:2015fha}.
In this work, we shall take
the factorization approximation as a working hypothesis rather
than a first-principles starting point as factorization has not been proven for three-body $B$ decays. That is, we
shall work in the phenomenological factorization model rather than
in the established QCD-inspired theories.

The layout of the present paper is as follows. In Sec. II we discuss resonant and nonresonant contributions to three-body $B$ decays.  The predicted rates for penguin-dominated $B\to VP$ modes are generally too small compared to experiment. We add power corrections induced by penguin annihilation to these modes to render a better agreement with the data. Sec. III is devoted to direct \CP violation. We consider inclusive and regional \CP asymmetries arising from both resonant and nonresonant mechanisms. The effect of final-state rescattering is discussed. Comparison of our work with others available in the literature is made in Sec/ IV. Sec. V contains our
conclusions.

%%%%%%%%%%%%%%%%%%%%%%%%%%%%%%
\section{Three-body decays}
%%%%%%%%%%%%%%%%%%%%%%%%%%%%%%

Many three-body $B$ decays have been observed with branching fractions of order $10^{-5}$ for penguin-dominated  $B\to K\pi\pi, KKK$ decays and
of order $10^{-6}$ for tree-dominated $B\to \pi\pi\pi, KK\pi$.  The charmless three-body channels that have been measured are \cite{PDG}:
\be
&& B^-\to \pi^+\pi^-\pi^-, K^-\pi^+\pi^-, \ov K^0\pi^-\pi^0, K^+K^-\pi^-, K^+K^-K^-, K^-\pi^0\pi^0, K^-K_SK_S, K_S\pi^-\pi^0, \non \\
&& \ov B^0\to \pi^+\pi^-\pi^0, \ov K^0\pi^+\pi^-, K^-\pi^+\pi^0, K^+K^-\pi^0, K^0K^-\pi^+, \ov K^0K^+\pi^-,K^+K^-\ov K^0, K_SK_SK_S,\non \\
&& \ov B_s^0\to K^0\pi^+\pi^-, K^0 K^+K^-,\ov K^0K^-\pi^+,K^0K^+\pi^-.
\en
In $B^-$ and $\ov B^0$ three-body decays, the $b\to sq\bar q$ penguin transitions
contribute to the final states with odd number of kaons, namely,
$KKK$ and $K\pi\pi$, while $b\to uq\bar q$ tree and $b\to dq\bar
q$ penguin transitions contribute to final states with even number
of kaons, e.g. $KK\pi$ and $\pi\pi\pi$.
For $\ov B_s^0$ three-body decays, the situation is the other way around.

\begin{figure}[t]
\centering
    \includegraphics[scale=0.5]{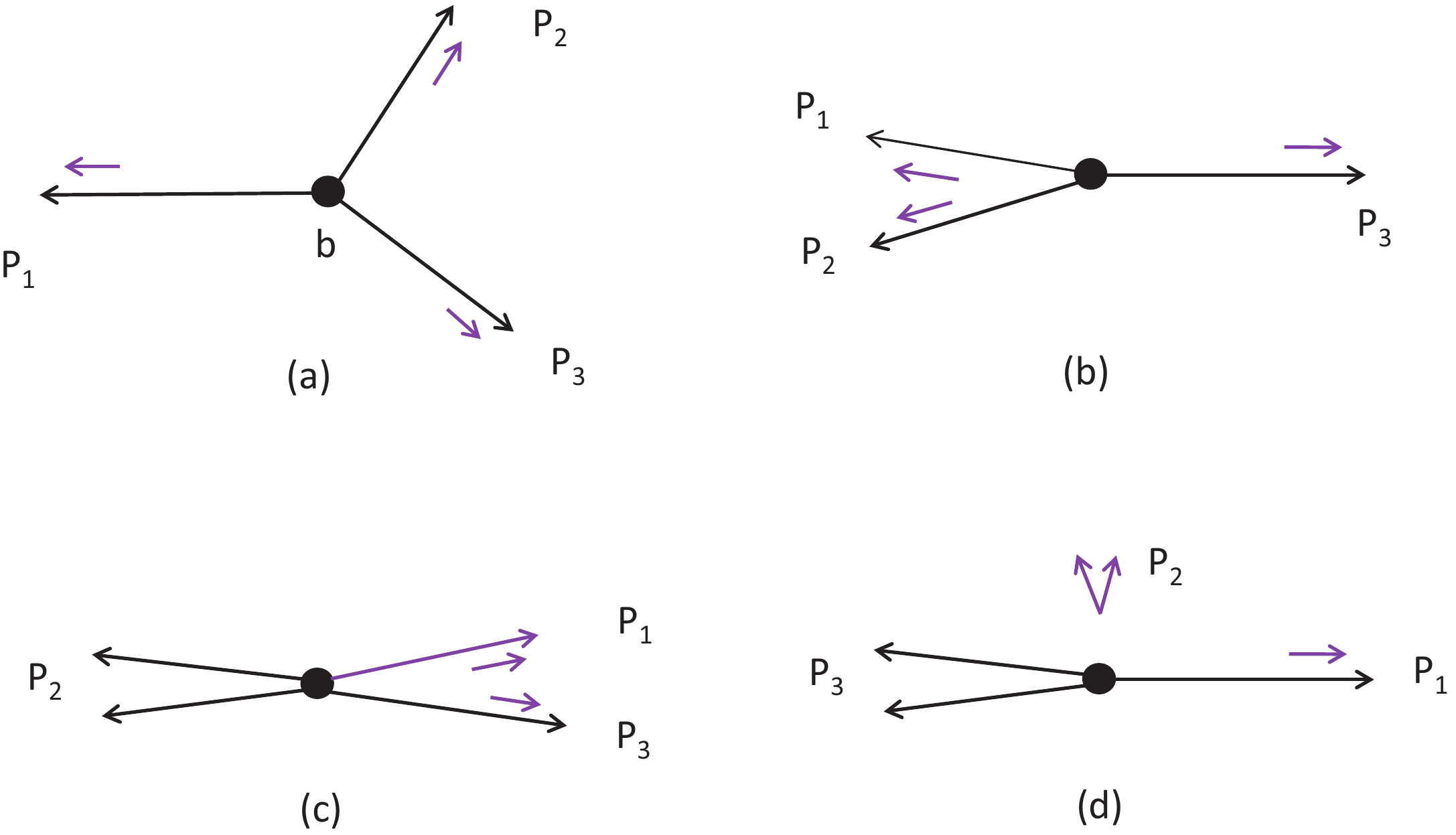}
\caption{Possible configurations of three-body $B\to P_1P_2P_3$ decays where the black lines with arrows denote the momenta of the three energetic quarks $q_1q_2\bar q_3$ produced in the $b$-quark decay  and the pink lines with arrows denote the momenta of the spectator quark and the quark-antiquark pair:  (a) all three produced mesons are moving energetically, (b) two of the energetic mesons, say $P_1$ and $P_2$, are moving collinearly to each other, recoiling against $P_3$, (c) $P_2$ is formed from $q_1\bar q_3$ or $q_2\bar q_3$, while $P_1$ contains the spectator quark (denoted by the longer pink line) which becomes hard after being  kicked by a hard gluon, and (d) is similar to (c) except that $P_2$ is soft.}
\label{fig:Config}
\end{figure}

\begin{figure}[t]
\centering
 {
      \includegraphics[scale=0.42]{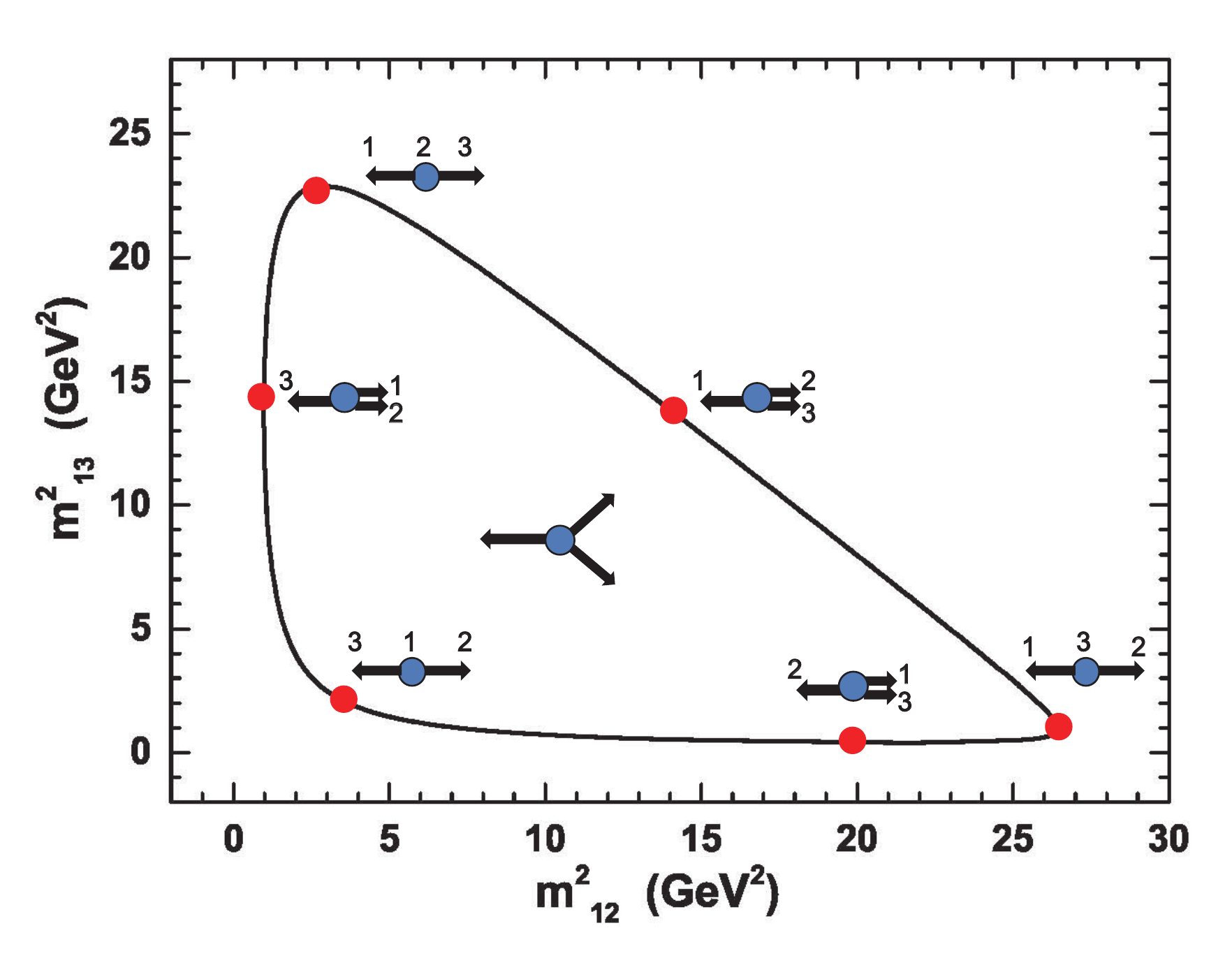}
}{
     \includegraphics[scale=0.3]{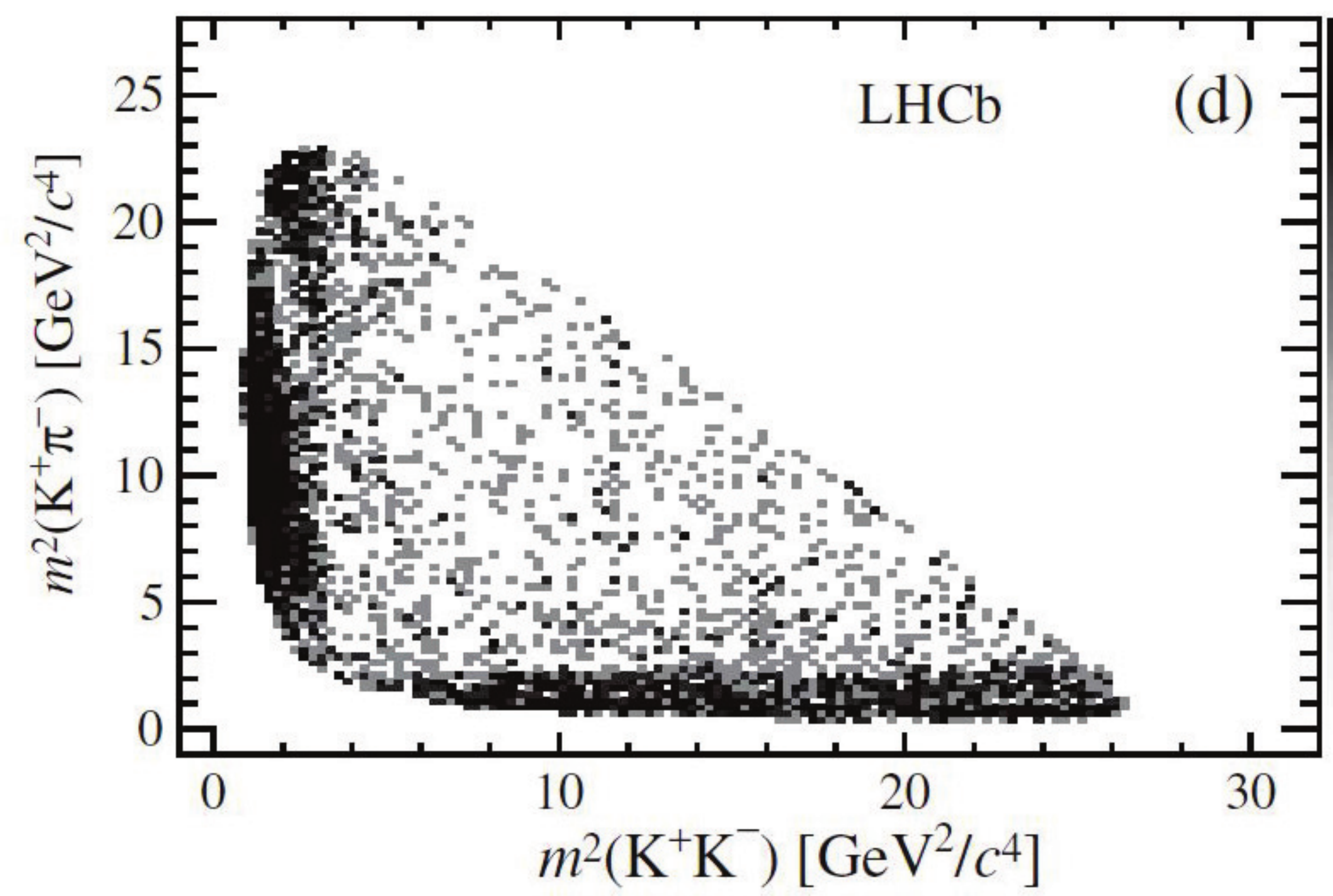}
}
\caption{(a) Location of various physical configurations depicted in Fig. \ref{fig:Config} within the Dalitz plot of $B^-\to K^+(p_1)K^-(p_2)\pi^-(p_3)$ and (b) the measured Dalitz plot distribution taken from \cite{LHCb:2014}. }
\label{fig:Dalitz}
\end{figure}

Consider the 3-body decays $B\to P_1P_2P_3$. The $b$ quark decays into three energetic quarks, $q_1q_2\bar q_3$.
There exist four possible physical configurations depicted in Fig. \ref{fig:Config}:
(a) all three produced mesons are moving energetically, (b) two of the energetic mesons, say $P_1$ and $P_2$, are moving collinearly to each other, (c) $P_3$ is formed from $q_1\bar q_3$ or $q_2\bar q_3$, while $P_2$ contains the spectator quark  which becomes hard after being  kicked by a hard gluon, and (d) is the same as (c) except that $P_2$ is soft. Configurations (b) and (c) mimic quasi-two-body decays. In the Dalitz plot of Fig. \ref{fig:Dalitz}, configuration (a) appears in the central region, while configurations (b)--(d) manifest along the edges of the Dalitz plot. The two mesons $P_1$ and $P_2$ in (b) move collinearly, recoiling against $P_3$. Hence, the invariant mass squared $m_{12}^2$ is minimal, while the momentum $p_3$ of $P_3$ is maximal. Likewise, configuration (c) has minimal $m_{13}^2$.
Resonances show up in configurations (b) and (c), corresponding to quasi-two-particle decays. Therefore, the Dalitz plot for three-body $B$ decays can be divided into several sub-regions with distinct kinematics and factorization properties, which have been investigated in \cite{Krankl:2015fha}. Especially, the regions containing the configuration (b) or (c) can be described in terms of two-meson distribution amplitudes and $B\to P_1P_2$ form factors \cite{Faller:2013dwa,Hambrock:2015aor,Dyk:Honnef}.

With the advent of heavy quark effective theory,
nonleptonic $B$ decays can be analyzed systematically within the QCD framework.
There are three popular approaches available in this regard: QCDF, pQCD and SCET.
Theories of hadronic $B$ decays are based on the ``factorization theorem"
under which the short-distance contributions to the decay amplitudes can be
separated from the process-independent long-distance parts. In the QCDF approach, nonfactorizable contributions to the hadronic
matrix elements can be absorbed into the effective parameters $a_i$
\begin{eqnarray}
A(B\to M_1M_2)={G_F\over \sqrt{2}}\sum \lambda_i a_i(M_1M_2)\langle M_1M_2|O_i|B\rangle_{\rm fact},
\end{eqnarray}
where $a_i$ are basically the Wilson coefficients
in conjunction with short-distance nonfactorizable corrections such as vertex,
penguin  corrections and hard spectator interactions, and
$\langle M_1M_2|O_i|B\rangle_{\rm fact}$ is the matrix element evaluated under
the factorization approximation.
Since power corrections of order $\Lambda_{\rm QCD}/m_b$ are suppressed in the heavy
quark limit, nonfactorizable corrections to nonleptonic decays are calculable.
In the limits of $m_b\to\infty$ and $\alpha_s\to 0$, naive factorization is
recovered in both QCDF and  pQCD approaches.

Unlike hadronic 2-body $B$ decays, established theories such as QCDF, pQCD and SCET are still not available for three-body decays, though attempts along the framework of pQCD and QCDF have been made in the past \cite{Chen:2002th,Wang:2014ira,Krankl:2015fha}. This is mainly because the aforementioned factorization theorem has not been proven for three-body decays. Hence, we follow \cite{Cheng:2013dua,CCS:nonres} to take the factorization approximation as a working hypothesis rather than a first-principles starting point.

One of the salient features of three-body $B$ decays is the large nonresonant fraction in penguin-dominated $B$ decay modes, recalling that the nonresonant signal in charm decays is very small,
less than 10\% \cite{PDG}. Many of the
charmless $B$ to three-body decay modes have been measured at $B$
factories and studied using the Dalitz-plot analysis. The
measured fractions and the corresponding branching fractions of
nonresonant components are
summarized in Table \ref{tab:BRexpt}.
We see that the nonresonant fraction is about $\sim (70-90)\%$
in $B\to K\!K\!K$ decays, $\sim (17- 40)\%$  in $B\to K\!\pi\pi$ decays, and $\sim$ 35\% in the $B\to\pi\pi\pi$
decay. Moreover, we have the hierarchy pattern
\be
\B(B\to KKK)_{\rm NR}> \B(B\to K\pi\pi)_{\rm NR} > \B(B\to \pi\pi\pi)_{\rm NR}.
\en
Hence, the nonresonant contributions play an essential
role in penguin-dominated $B$ decays. This is not
unexpected because the energy release scale in weak $B$
decays is of order 5 GeV, whereas the major resonances lie in the
energy region of 0.77 to 1.6 GeV. Consequently, it is likely that
three-body $B$ decays will receive sizable nonresonant contributions.
It is important to
understand and identify the underlying mechanism for nonresonant
decays.

%%%%%%%%%%%%%%%%%%%%%%%%%%%%%%%%%%%%%%%%%%%%%%
\begin{table}[t]
\caption{The fractions and branching fractions of nonresonant components of various charmless three-body decays
of $B$ mesons.  } \label{tab:BRexpt}
 \footnotesize{
\begin{ruledtabular}

\begin{tabular}{l c c c c c}
 &  \multicolumn{2}{c}{BaBar}
 &   \multicolumn{2}{c}{Belle} \\ \cline{2-3} \cline{4-5}
\raisebox{2.0ex}[0cm][0cm]{Decay}  &  $\B_{\rm NR}(10^{-6})$ & NR fraction(\%) & $\B_{\rm NR}(10^{-6})$ &  NR fraction(\%) & \raisebox{2.0ex}[0cm][0cm]{Reference} \\ \hline
 $B^-\to K^+K^-K^-$ & $22.8\pm2.7\pm7.6$&  $68.3\pm8.1\pm22.8$  & $24.0\pm1.5\pm1.5$ & $78.4\pm5.8\pm7.7$  &\cite{BaBarKKK,BelleKpKpKm} \\
 $B^-\to K^-K_SK_S$  &  $19.8\pm3.7\pm2.5$ & $\sim$196   &  & & \cite{BaBarKKK}  \\
 $\ov B^0\to K^+K^-\ov K^0$ & $33\pm5\pm9$ & $\sim$130 & &     & \cite{BaBarKKK} \\
 $\ov B^0\to K_SK_SK_S$ & $13.3^{+2.2}_{-2.3}\pm2.2$ & $\sim$215  & &  & \cite{BaBarKsKsKs} \\
 $B^-\to K^-\pi^+\pi^-$ & $9.3\pm1.0^{+6.9}_{-1.7}$ & $17.1\pm1.7^{+12.4}_{-~1.8}$  & $16.9\pm1.3^{+1.7}_{-1.6}$ & $34.0\pm2.2^{+2.1}_{-1.8}$  & \cite{BaBar:Kmpippim,Belle:Kmpippim} \\
 $ \ov B^0\to \ov K^0\pi^+\pi^-$  & $11.1^{+2.5}_{-1.0}\pm0.9$ & $22.1^{+2.8}_{-2.0}\pm2.2$  & $19.9\pm2.5^{+1.7}_{-2.0}$ & $41.9\pm5.1^{+1.5}_{-2.6}$  &\cite{BaBarK0pippim,BelleK0pipi} \\
 $\ov B^0\to K^-\pi^+\pi^0$ &  $7.6\pm0.5\pm1.0$ & $19.7\pm1.4\pm3.3$  &  $5.7^{+2.7+0.5}_{-2.5-0.4}$ &  $<25.7$ & \cite{BaBarKppimpi0,BelleKppimpi0} \\
 $B^- \to\pi^+\pi^-\pi^-$ &  $5.3\pm0.7^{+1.3}_{-0.8}$ & $34.9\pm4.2^{+8.0}_{-4.5}$ & & & \cite{BaBarpipipi} \\
\end{tabular}
\end{ruledtabular} }
 \end{table}
%%%%%%%%%%%%%%%%%%%%%%%%%%%%%

It has been argued in \cite{Krankl:2015fha} that the Dalitz plot is completely dominated by the edges as the amplitude at the center is both power- and $\alpha_s$-suppressed with respect to the one at the edge. As a result, three-body decays become quasi two-body ones. Nevertheless, this argument is not inconsistent with the experimental observation of dominant nonresonant background in penguin-dominated 3-body decays because the nonresonant background exists in the whole phase space. That is, the vast phase space of charmless three-body $B$ decays is populated by nonresonant components.

The explicit expressions of  factorizable amplitudes of charmless $B\to P_1P_2P_3$ decays can be found in \cite{Cheng:2013dua,CCS:nonres}.
There are three distinct
factorizable terms: (i) the current-induced process with a meson
emission, $\la B\to P_1\ra\times \la 0\to P_2P_3\ra$,
(ii) the transition process,  $\la B\to P_1P_2\ra\times \la
0\to P_3\ra$, and (iii) the annihilation process $\la
B\to 0\ra\times \la 0\to P_1P_2P_3\ra$, where $\la A\to
B\ra$ denotes a $A\to B$ transition matrix element. There are two different kinds of mechanisms for the production of a meson pair. In $\la 0\to P_2P_3\ra$, the meson pair is produced from the vacuum through a current, whereas in $\la B\to P_1P_2\ra$ the meson pair is produced through a current that induces the transition from the $B$ meson. Hence, we call these as current-induced and transition mechanisms, respectively.~\footnote{Note that the terminology concerning current-induced and transition mechanisms in this work is different to those in our previous publications \cite{Cheng:2013dua,CCS:nonres}.}
While the latter process is produced at the $b\to u$ tree level, the former one is induced at the $b\to s$ or $b\to d$ penguin level. Schematically, the decay amplitude is the coherent sum of resonant contributions together with the nonresonant background
\be
A=\sum_R A_R+A_{\rm NR}.
\en
In the following,
we will discuss these two contributions separately.

\subsection{Nonresonant background}

Consider the transition process induced by the $b\to u$ current.
The nonresonant contribution to the three-body matrix
element $\la P_1 P_2|(\bar u b)_{V-A}|B\ra$ has
the general expression~\cite{LLW}
\be \label{eq:romegah}
 \la P_1 (p_1) P_2(p_2)|(\bar u b)_{V-A}|B\ra^{\rm NR}
 &=&i r
 (p_B-p_1-p_2)_\mu+i\omega_+(p_2+p_1)_\mu+i\omega_-(p_2-p_1)_\mu
 \non\\
 &&+h\,\epsilon_{\mu\nu\alpha\beta}p_B^\nu (p_2+p_1)^\alpha
 (p_2-p_1)^\beta,
\en
where $(\bar q_1q_2)_{V-A}=\bar q_1\gamma_\mu(1-\gamma_5)q_2$.
The form factors $r$,
$\omega_\pm$ and $h$ can be evaluated in the framework of heavy meson chiral perturbation theory (HMChPT) \cite{LLW}. Consequently, the nonresonant amplitude induced by the transition process reads
\be \label{eq:AHMChPT}
 A_{\rm transition}^{\rm HMChPT} &\equiv&\la P_3(p_3)|(\bar q
 u)_{V-A}|0\ra \la   P_1 (p_1) P_2(p_2)|(\bar u b)_{V-A}|B\ra^{\rm NR} \non\\
 &=& -\frac{f_{P_3}}{2}\left[2 m_3^2 r+(m_B^2-s_{12}-m_3^2) \omega_+
 +(s_{23}-s_{13}-m_2^2+m_1^2) \omega_-\right].
\en
However, as pointed out in \cite{Cheng:2013dua,CCS:nonres}, the predicted nonresonant rates based on HMChPT are unexpectedly too large for tree-dominated decays. For example, the branching fractions of nonresonant $B^-\to \pi^+\pi^-\pi^-$ and $B^-\to K^+K^-\pi^-$ are found to be of order
$75\times 10^{-6}$ and $33\times 10^{-6}$, respectively, which are one order of magnitude larger than the corresponding measured total branching fractions of $15.2\times 10^{-6}$ and $5.0\times 10^{-6}$ (see Table \ref{tab:Kpipi} below).
The issue has to do with the applicability of HMChPT. In order to
apply this approach, two of the final-state pseudoscalars in $B\to
P_1P_2$ transition have to be soft; their momenta should be smaller than the chiral symmetry breaking
scale of order 1 GeV.
Therefore, it is not justified to apply chiral and heavy quark
symmetries to a certain kinematic region and then generalize it to
the region beyond its validity.
Following \cite{CCS:nonres}, we shall assume the momentum dependence of nonresonant amplitudes in an exponential
form, namely,
\be \label{eq:ADalitz}
  A_{\rm transition}=A_{\rm transition}^{\rm
  HMChPT}\,e^{-\alpha_{_{\rm NR}}
p_B\cdot(p_1+p_2)}e^{i\phi_{12}},
\en
so that the HMChPT results are recovered in the soft meson limit of 
$p_1,~p_2\to 0$.  This is similar to the empirical parametrization of the non-resonant
amplitudes adopted in the BaBar and Belle analyses \cite{BelleKpKpKm,BaBar:NR}
 \be \label{eq:ANR}
A_{\rm NR}=c_{12}e^{i\phi_{12}}e^{-\alpha
s_{12}}+c_{13}e^{i\phi_{13}}e^{-\alpha
s_{13}}+c_{23}e^{i\phi_{23}}e^{-\alpha s_{23}}.
 \en
We
shall use the tree-dominated $B^-\to\pi^+\pi^-\pi^-$ decay data to
fix the unknown parameter $\alpha_{_{\rm NR}}$ as its nonresonant component is predominated by the transition process. Hence, the measurement of nonresonant contributions to $B^-\to\pi^+\pi^-\pi^-$ provides an ideal place to constrain the
parameter $\alpha_{_{\rm NR}}$, which turns out to be \cite{Cheng:2013dua}
\be \label{eq:alphaNR}
 \alpha_{_{\rm NR}}=0.081^{+0.015}_{-0.009}\,{\rm GeV}^{-2}.
\en
The phase $\phi_{12}$ of the nonresonant amplitude will be set to zero for simplicity.

Note that $A_{\rm transition}^{\rm HMChPT}$ receives nonresonant contributions from the whole Dalitz plot, including the central regions and regions near and along the edge. Since $p_B\cdot(p_1+p_2)={1\over 2}(m_B^2-m_3^2+s_{12})$, it is obvious that the nonresonant signal $A_{\rm transition}$ arises mainly from the small invariant mass region of $s_{12}$.

For penguin-dominated decays $B\to KKK$ and $B\to K\pi\pi$,
the nonresonant background induced from the $b\to u$ transition process yields $\B(B^-\to K^+K^-K^-)^{\rm NR}\sim 1.1\times 10^{-6}$ and $\B(B^-\to K^+\pi^+\pi^-)^{\rm NR}\sim 0.8\times 10^{-6}$, which are too small compared to experiment (see Table \ref{tab:Kpipi}). This is ascribed to the large CKM suppression $|V_{ub}V^*_{us}|\ll |V_{cb}V^*_{cs}|\approx |V_{tb}V^*_{ts}|$ associated with the $b\to u$ tree transition relative to the $b\to s$ penguin process.
This implies that the two-body
matrix element of scalar densities e.g. $\langle K\overline K|\bar ss|0\rangle$ induced from the penguin diagram should
have a large nonresonant component. The explicit expression of the nonresonant component of $\langle K\overline K|\bar ss|0\rangle$ will be shown in Eq. (\ref{eq:KKssme}) below.

For the nonresonant contributions to the 2-body matrix elements $\la P_1P_2|\bar q\gamma_\mu q'|0\ra$ and $\la P_1P_2|\bar qq'|0\ra$, we shall use the measured  kaon electromagnetic form factors to extract $\la K\ov K|\bar q\gamma_\mu q'|0\ra^{\rm NR}$ and $\la K\ov K|\bar ss|0\ra^{\rm NR}$ first and then apply SU(3) symmetry to relate them to other 2-body matrix elements \cite{CCS:nonres}.

\subsection{Resonant contributions}
In the experimental analysis of three-body decays, the resonant amplitude associated with the intermediate resonance $R$ takes the form \cite{Bevan:2014iga}
\be
A_R=F_P\times F_R\times T_R\times W_R,
\en
where $T_R$ is usually described by a relativistic Breit-Wigner parametrization, $W_R$ accounts for the angular distribution of the decay, $F_P$ and $F_R$ are the transition form factors of the parent particle and resonance, respectively (see e.g. \cite{Bevan:2014iga} for details).

In general, vector meson and scalar resonances contribute to the two-body matrix elements $\langle P_1P_2|V_\mu|
0\rangle$ and $\langle
P_1P_2|S|0\rangle$, respectively. The
intermediate vector meson contributions to three-body decays are
identified through the vector current, while the scalar meson
resonances are mainly associated with the scalar density. Both scalar and vector resonances can contribute to
the three-body matrix element $\la P_1P_2|J_\mu|B\ra$.
Effects of intermediate resonances are described as a coherent sum of Breit-Wigner
expressions. More precisely,
\footnote{Strictly speaking, for the $f_0(980)$ and $a_0(980)$ we should use the Flatt\'e parametrization \cite{Flatte} to account for the threshold effect, though in practice we find that numerically it makes no significant difference from the use of the  Breit-Wigner propagator.}
\begin{eqnarray}
  \la P_1(p_1)P_2(p_2)|(\bar qb)_\vma|B\ra^R &=& \sum_i\langle P_1P_2|V_i\rangle
 {1\over s_{12}-m_{V_i}^2+im_{V_i}\Gamma_{V_i}}\langle V_i|(\bar qb)_\vma|B\rangle \non \\
&+& \sum_i\langle P_1P_2|S_i\rangle
 {-1\over s_{12}-m_{S_i}^2+im_{S_i}\Gamma_{S_i}}\langle S_i|(\bar qb)_\vma|B\rangle, \non \\
  \langle P_1P_2|\bar q_1\gamma_\mu q_2|0\rangle^R &=& \sum_i\langle P_1P_2|V_i\rangle
 {1\over s_{12}-m_{V_i}^2+im_{V_i}\Gamma_{V_i}}\langle V_i|\bar q_1\gamma_\mu q_2|0\rangle, \nonumber \\
 &+& \sum_i\langle P_1P_2|S_i\rangle
 {-1\over s_{12}-m_{S_i}^2+im_{S_i}\Gamma_{S_i}}\langle S_i|\bar q_1\gamma_\mu q_2|0\rangle,\non \\
 \langle P_1P_2|\bar q_1q_2|0\rangle^R &=& \sum_i\langle P_1P_2|S_i\rangle
 {-1\over s_{12}-m_{S_i}^2+im_{S_i}\Gamma_{S_i}}\langle S_i|\bar q_1q_2|0\rangle,
 \end{eqnarray}
where $V_i=\phi,\rho,\omega,\cdots$ and
$S_i=f_0(980),f_0(1370),f_0(1500),\cdots$ for $P_1P_2=\pi^+\pi^-$, and
$V_i=K^*(892),K^*(1410),K^*(1680),\cdots$ and
$S_i=K^*_0(1430),\cdots$ for $P_1P_2=K^\pm\pi^\mp$. In general, the decay widths $\Gamma_{V_i}$ and $\Gamma_{S_i}$ are  energy dependent. For $f_0(500)$ and $K_0^*(800)$, they are too broad to use the Breit-Wigner formulism.

Notice that the two-body matrix element $\la P_1P_2|V_\mu|0\ra$ can also
receive contributions from scalar resonances when $q_1\neq q_2$.  For example, both $K^*$ and $K_0^*(1430)$ contribute to the matrix
element $\la K^-\pi^+|\bar s\gamma_\mu d|0\ra$ given by
\be \label{eq:m.e.pole2}
\la K^-(p_1)\pi^+(p_2)|\bar s\gamma_\mu d|0\ra^R &=& \sum_i
{g^{K^*_i\to K^-\pi^+}\over
s_{12}-m_{K^*_i}^2+im_{K^*_i}\Gamma_{K^*_i}}\sum_{\rm
pol}\vp^*\cdot
(p_1-p_2)\la K^*_i|\bar s\gamma_\mu d|0\ra \non \\
&-& \sum_i{g^{{K^*_{0i}}\to K^-\pi^+}\over
s_{12}- m_{K^*_{0i}}^2+im_{K^*_{0i}}\Gamma_{K^*_{0i}}}\la
K^*_{0i}|\bar s\gamma_\mu d|0\ra,
\en
with $K_i^*=K^*(892), K^*(1410),K^*(1680),\cdots$, and $K_{0i}^*=K_0^*(800),K_0^*(1430),\cdots$.

\subsection{Nonresonant contribution from matrix element of scalar density}

Consider the nonresonant amplitude in the penguin-dominated $B^-\to K^+K^-K^-$ decay. In addition to the $b\to u$ tree transition which yields a rather small nonresonant fraction, we need to consider the
nonresonant amplitudes indcued from the $b\to s$ penguin transition
\begin{eqnarray}
 A_1 &=& \langle K {}^-(p_1)|(\bar s b)_{V-A}|B {}^-\rangle
  \langle K^+(p_2) K^-(p_3)|(\bar qq)_{V-A}|0\rangle, \nonumber \\
 A_2 &=& \langle K {}^-(p_1)|\bar s b|B {}^-\rangle
       \langle K^+(p_2) K^-(p_3)|\bar s s|0\rangle,
\end{eqnarray}
for $q=u,d,s$. The two-kaon matrix element created from the vacuum can be expressed in terms of time-like kaon current form factors as
 \be \label{eq:KKweakff}
 \la K^+(p_{K^+}) K^-(p_{K^-})|\bar q\gamma_\mu q|0\ra
 &=& (p_{K^+}-p_{K^-})_\mu F^{K^+K^-}_q,
 \non\\
 \la K^0(p_{K^0}) \ov K^0(p_{\bar K^0})|\bar q\gamma_\mu q|0\ra
 &=& (p_{K^0}-p_{\bar K^0})_\mu F^{K^0\bar K^0}_q.
 \en
The weak vector form factors $F^{K^+K^-}_q$ and $F^{K^0\bar
K^0}_q$ can be related to the kaon e.m. form
factors $F^{K^+K^-}_{\rm em}$ and $F^{K^0\bar K^0}_{\rm em}$ for the
charged and neutral kaons, respectively. As shown in \cite{CCS:nonres}, the nonresonant components of $F^{K^+K^-}_q$ read
 \be
 F^{K^+K^-}_{u,N\!R}=\frac{1}{3}(3F_{N\!R}-F'_{N\!R}),
 \qquad F^{K^+K^-}_{d,N\!R}=0, \qquad
 F^{K^+K^-}_{s,N\!R}=-\frac{1}{3}(3 F_{N\!R}+2F'_{N\!R}),
 \label{eq:FKKisospin}
 \en
where the nonresonant terms $F_{N\!R}$ and $F'_{N\!R}$ can be
parameterized as
 \be
 F^{(\prime)}_{NR}(s_{23})=\left(\frac{x^{(\prime)}_1}{s_{23}}
 +\frac{x^{(\prime)}_2}{s_{23}^2}\right)
 \left[\ln\left(\frac{s_{23}}{\tilde\Lambda^2}\right)\right]^{-1},
 \en
with $\tilde\Lambda\approx 0.3$ GeV.  The unknown parameters
$x_i$ and $x'_i$ are fitted from the kaon e.m. data, see \cite{DKK} for details.

The nonresonant component of the matrix element of scalar density is given by \cite{CCS:nonres}
\footnote{Matrix elements of scalar densities (or scalar form factors) have also been studied in \cite{Doring:2013wka} within the framework of unitarized chiral perturbation theory and dispersion relations. However, the main focus there is on resonant contributions.}
\be \label{eq:KKssme}
 \la K^+(p_2) K^-(p_3)|\bar s s|0\ra^{\rm NR}
=\frac{v}{3}(3 F_{NR}+2F'_{NR})+\sigma_{_{\rm NR}}
 e^{-\alpha s_{23}}.
\en
with
 \be \label{eq:v}
 v=\frac{m_{K^+}^2}{m_u+m_s}=\frac{m_K^2-m_\pi^2}{m_s-m_d}.
 \en
From the measured $\overline B^0\to K_SK_SK_S$ rate and the $K^+K^-$ mass
spectrum measured in $\ov B^0\to K^+K^-K_S$, the nonresonant $\sigma_{_{\rm NR}}$ term  can be constrained to be \cite{CCS:nonres}
\be \label{eq:sigma}
  \sigma_{_{\rm NR}}= e^{i\pi/4}\left(3.39^{+0.18}_{-0.21}\right)\,{\rm GeV}.
\en
For the
parameter $\alpha$ appearing in Eq. (\ref{eq:KKssme}), we will use the experimental measurement
$\alpha=(0.14\pm0.02)\,{\rm GeV}^{-2}$ \cite{BaBarKpKmK0}.
Numerically, the nonresonant signal is governed by the $\sigma_{_{\rm NR}}$ component of the matrix element of scalar density. Owing to the exponential suppression factor $e^{-\alpha \,s_{ij}}$ in Eq. (\ref{eq:KKssme}), the nonresonant contribution manifests in the low invariant mass regions.

\subsection{Branching fractions}
For numerical calculations we follow \cite{Cheng:2013dua} for the input parameters except the CKM matrix elements, which we will use the updated Wolfenstein parameters
$A=0.8227$, $\lambda=0.22543$, $\bar \rho=0.1504$ and $\bar
\eta=0.3540$ \cite{CKMfitter}. The corresponding CKM angles are
$\sin2\beta=0.710\pm0.011$ and
$\gamma=(67.01^{+0.88}_{-1.99})^\circ$ \cite{CKMfitter}. In Table \ref{tab:Kpipi} we present updated branching fractions of resonant and
nonresonant components in $B^-\to K^+K^-K^-, K^-\pi^+\pi^-, K^+K^-\pi^-$ and $\pi^-\pi^+\pi^-$ decays.

\subsubsection{$B^-\to K^+K^-K^-$}

As shown before in \cite{Cheng:2013dua}, the calculated $B^-\to K^-\phi\to K^-K^+K^-$ rate in the factorization approach is smaller than experiment. In the QCD factorization approach, this rate deficit problem calls
for the $1/m_b$ power corrections from penguin annihilation. In this approach, it amounts to replacing the penguin contribution characterized by $a_4^p\to a_4^p+\beta_3^p$, where $p=u,c$ and $\beta_3$ is the annihilation contribution induced mainly from $(S-P)(S+P)$ operators \cite{BN}. For our purpose we will use
\be
\beta_3^u[K\!\phi]=\beta_3^c[K\!\phi]=-0.0085+0.0088i\,.
\en
This power correction $\beta_3^p[K\!\phi]$ is calculated in \cite{CC:Bud} for the quasi-two-body decay $B^-\to K^-\phi$. In principle, it should be computed in the 3-body decay $B^-\to K^+K^-K^-$ with $m(K^+K^-)_{\rm low}$ peaked at the $\phi$ mass in QCDF. We will assume that $\beta_3^p[K\!\phi]$ calculated in either way is similar.

From Table \ref{tab:Kpipi} it is clear that the predicted rates for
the nonresonant component and for the total branching fraction of $B^-\to K^+K^-K^-$ are consistent with both BaBar and Belle within errors.

\begin{table}[!]
\footnotesize{
\caption{Branching fractions (in units of $10^{-6}$) of resonant and
nonresonant (NR) contributions to $B^-\to \pi^-\pi^+\pi^-, K^-\pi^+\pi^-, K^+K^-\pi^-, K^+K^-K^-$.  Note that  the BaBar result for
$K_0^{*0}(1430)\pi^-$  in \cite{BaBar:Kmpippim} is their
absolute one. We have converted them into the product branching
fractions, namely, $\B(B\to Rh)\times \B(R\to hh)$.
The nonresonant background in $B^-\to\pi^+\pi^-\pi^-$ is used as an input to
fix the parameter $\alpha_{_{\rm NR}}$ defined in Eq.
(\ref{eq:ADalitz}).
Theoretical errors correspond to the uncertainties in (i)
$\alpha_{_{\rm NR}}$, (ii) $F^{B\pi}_0$, $\sigma_{_{\rm
NR}}$ and $m_s(\mu)=(90\pm 20) $MeV at $\mu=2.1$ GeV,  and (iii) $\gamma=(67.01^{+0.88}_{-1.99})^\circ$.
}
\begin{ruledtabular}
\begin{tabular}{l l l l} \label{tab:Kpipi}
 $B^-\to K^+K^-K^-$
   \\
 Decay mode~~
   & BaBar \cite{BaBarKKK}
   & Belle \cite{BelleKpKpKm}
   & Theory
   \\
 \hline
$\phi K^-$
   & $4.48\pm0.22^{+0.33}_{-0.24}$
   &$4.72\pm0.45\pm0.35^{+0.39}_{-0.22}$
   & $4.4^{+0.0+0.8+0.0}_{-0.0-0.7-0.0}$  %
   \\
$f_0(980)K^-$
   & $9.4\pm1.6\pm2.8$
   & $<2.9$
   & $11.2^{+0.0+2.7+0.0}_{-0.0-2.1-0.0}$%%
   \\
$f_0(1500)K^-$
   & $0.74\pm0.18\pm0.52$
   &
   & $0.63^{+0.0+0.11+0.0}_{-0.0-0.10-0.0}$%%
   \\
$f_0(1710)K^-$
   & $1.12\pm0.25\pm0.50$
   &
   & $1.2^{+0+0.2+0}_{-0-0.2-0}$%%
   \\
$f'_2(1525)K^-$
   & $0.69\pm0.16\pm0.13$
   &
   &
   \\
NR
   & $22.8\pm2.7\pm7.6$
   & $24.0\pm1.5\pm1.8^{+1.9}_{-5.7}$
   & $21.1^{+0.8+7.2+0.1}_{-1.1-5.7-0.1}$ %
   \\
\hline
Total
   & $33.4\pm0.5\pm0.9$
   & $30.6\pm1.2\pm2.3$
   &  $28.8^{+0.5+7.9+0.1}_{-0.6-6.4-0.1}$%
   \\ \hline \hline
   $B^-\to K^-\pi^+\pi^-$ \\
Decay mode~~
  & BaBar \cite{BaBar:Kmpippim}
  & Belle \cite{Belle:Kmpippim}
  & Theory
\\
\hline
$\overline K^{*0}\pi^-$
  & $7.2\pm0.4\pm0.7^{+0.3}_{-0.5}$
  & $6.45\pm0.43\pm0.48^{+0.25}_{-0.35}$
  &  $8.4^{+0.0+2.1+0.0}_{-0.0-1.9-0.0}$%%
  \\
$\overline K^{*0}_0(1430)\pi^-$
  & $19.8\pm0.7\pm1.7^{+5.6}_{-0.9}\pm3.2$ \footnotemark[1]
  &$32.0\pm1.0\pm2.4^{+1.1}_{-1.9}$
  & $11.5^{+0.0+3.3+0.0}_{-0.0-2.8-0.0}$ %%
  \\
$\rho^0K^-$ & $3.56\pm0.45\pm0.43^{+0.38}_{-0.15}$
  &$3.89\pm0.47\pm0.29^{+0.32}_{-0.29}$
  & $2.9^{+0.0+0.7+0.0}_{-0.0-0.2-0.0}$%%
  \\
$f_0(980)K^-$
  & $10.3\pm0.5\pm1.3^{+1.5}_{-0.4}$
  & $8.78\pm0.82\pm0.65^{+0.55}_{-1.64}$
  & $6.7^{+0.0+1.6+0.0}_{-0.0-1.3-0.0}$%%
  \\
NR
  & $9.3\pm1.0\pm1.2^{+6.7}_{-0.4}\pm1.2$
  & $16.9\pm1.3\pm1.3^{+1.1}_{-0.9}$
  & $15.7^{+0.0+8.1+0.0}_{-0.0-5.2-0.0}$%%
  \\
\hline
Total
  & $54.4\pm1.1\pm4.6$
  & $48.8\pm1.1\pm3.6$
  & $42.2^{+0.2+16.1+0.1}_{-0.1-10.7-0.1}$ %%
  \\
 \hline \hline
$B^-\to K^+K^-\pi^-$
\\
Decay mode~~
  & BaBar \cite{BaBarKpKmpim}
  & Belle \cite{Belle2004}
  & Theory
\\
\hline
$K^{*0}K^-$
  & &
  & $0.21^{+0.00+0.04+0.00}_{-0.00-0.04-0.00}$
  \\
$K^{*0}_0(1430)K^-$
  & &
  & $1.0^{+0.0+0.2+0.0}_{-0.0-0.2-0.0}$
  \\
$f_0(980)\pi^-$
  & &
  & $0.25^{+0.00+0.01+0.00}_{-0.00-0.01-0.00}$
  \\
NR
  & &
  & $2.9^{+0.7+0.6+0.0}_{-0.8-0.4-0.0}$
  \\
\hline
Total
  & $5.0\pm0.7$
  & $<13$
  & $5.2^{+0.8+1.0+0.0}_{-0.9-0.7-0.0}$
  \\
  \hline \hline
$B^-\to \pi^-\pi^+\pi^-$
\\
Decay mode~~
  & BaBar \cite{BaBarpipipi}
  &
  & Theory
\\
\hline
$\rho^0\pi^-$
  & $8.1\pm0.7\pm1.2^{+0.4}_{-1.1}$
  &
  &  $7.3^{+0.0+0.4+0.0}_{-0.0-0.4-0.0}$ %
  \\
$\rho^0(1450)\pi^-$
  & $1.4\pm0.4\pm0.4^{+0.3}_{-0.7}$
  \\
$f_0(1370)\pi^-$
  & $2.9\pm0.5\pm0.5^{+0.7}_{-0.5}$
  &
  & $1.7^{+0.0+0.0+0.0}_{-0.0-0.0-0.0}$
  \\
$f_0(980)\pi^-$
  & $<1.5$
  &
  & $0.2^{+0.0+0.0+0.0}_{-0.0-0.0-0.0}$%
  \\
NR
  & $5.3\pm0.7\pm0.6^{+1.1}_{-0.5}$
  &
  & input
  \\
\hline
Total
  & $15.2\pm0.6\pm1.2^{+0.4}_{-0.3}$
  &
  & $17.0^{+2.0+0.9+0.2}_{-2.3-0.7-0.2}$
  \\
\end{tabular}
\end{ruledtabular}
\footnotetext[1]{\scriptsize Recently BaBar has measured the 3-body decay $B^-\to K_S^0\pi^-\pi^0$ and obtained $\B(B^-\to \ov K_0^{*0}(1430)\pi^-)=(31.0\pm3.0\pm3.8^{+1.7}_{-1.6})\times 10^{-6}$ \cite{BaBar:Kstpi}.}
 }
\end{table}

\subsubsection{$B^-\to K^-\pi^+\pi^-$}

We first discuss resonant decays. From Table VI of \cite{Cheng:2013dua}, it is obvious that except for
$f_0(980)K$,  the predicted rates for penguin-dominated channels $K^{*}\pi$,
$K^{*}_0(1430)\pi$ and $\rho K$ in $B^-\to K^-\pi^+\pi^-$ within the factorization approach are substantially smaller than the data by a factor of 2 $\sim$ 5. To overcome this problem, we shall use the penguin-annihilation induced power corrections alculated in our previous work \cite{CC:Bud}. The results are
\be \label{eq:beta}
\beta_3^p[\ov K^{*0}\pi^-]=-0.032+0.022i, \qquad \beta_3^p[\rho^0 K^-]=0.004-0.047i,
\en
for $p=u,c$.
It is evident the discrepancy between theory and experiment for $\ov K^{*0}\pi^-$ and $\rho^0 K^-$ is greatly improved (see Table \ref{tab:Kpipi}).

As for the quasi-2-body mode $B^-\to\ov K^{*0}_0(1430)\pi^-$, BaBar has recently measured the 3-body decay $B^-\to K_S^0\pi^-\pi^0$ and obtained $\B(B^-\to \ov K_0^{*0}(1430)\pi^-\to K^-\pi^+\pi^-)=(31.0\pm3.0\pm3.8^{+1.6}_{-1.6})\times 10^{-6}$ \cite{BaBar:Kstpi}. This is in good agreement with the Belle's result $(32.0\pm1.0\pm2.4^{+1.1}_{-1.9})\times 10^{-6}$ \cite{Belle:Kmpippim}. Hence, the predicted rate by naive factorization is too small by a factor of 3. Indeed, this is still an unresolved puzzle even in both QCDF and pQCD approaches \cite{Cheng:scalar,CCY:SP}. Using $\B(K_0^*(1430)\to K\pi)=0.93$, we find
$\B(B^-\to \ov K_0^{*0}(1430)\pi^-)_{\rm expt}\sim 51\times 10^{-6}$, while QCDF predicts $(12.9^{+4.6}_{-3.7})\times 10^{-6}$ \cite{Cheng:scalar}.
This explains why our prediction of the total branching
fraction of $B^-\to K^-\pi^+\pi^-$ is smaller than both BaBar and Belle.

The nonresonant component of $B\to K\!K\!K$ is governed by the $K\ov K$ matrix element of scalar density  $\la K\ov K|\bar
ss|0\ra$. By the same token, the nonresonant contribution to the penguin-dominated $B\to K\pi\pi$ decays should be also
dominated by the $K\pi$ matrix element of scalar density, namely $\la K\pi|\bar sq|0\ra$. When the unknown two-body matrix elements  such as $\la K^- \pi^+|\bar sd|0\ra$ and $\la \ov K^0 \pi^-|\bar su|0\ra$, $\la K^- \pi^0|\bar su|0\ra$ and $\la \ov K^0 \pi^0|\bar sd|0\ra$ are related to $\la K^+K^-|\bar ss|0\ra$ via SU(3) symmetry, e.g.
\be \label{eq:Kpim.e.SU3}
 \la K^-(p_1) \pi^+(p_2)|\bar sd|0\ra^{N\!R}=\la K^+(p_1)K^-(p_2)|\bar
 ss|0\ra^{N\!R},
\en
we find too large nonresonant and total branching fractions, namely $\B(B^-\to K^-\pi^+\pi^-)_{\rm NR}\sim 29.7\times 10^{-6}$ and $\B(B^-\to K^-\pi^+\pi^-)_{\rm tot}\sim 68.5\times 10^{-6}$.  Furthermore, Eq. (\ref{eq:Kpim.e.SU3}) will lead to  negative asymmetries $\A_{C\!P}^{\rm incl}(B^-\to K^-\pi^+\pi^-)\sim -0.8\%$ and $\A_{C\!P}^{\rm resc}(B^-\to K^-\pi^+\pi^-)\sim -6.4\%$ which are wrong in sign when confronted with the data. To accommodate the rates, it is tempting to assume that $\la K^-\pi^+|\bar sd|0\ra$ becomes slightly smaller because of SU(3) breaking. However, the predicted \CP asymmetry is still not correct in sign. As argued in \cite{Cheng:2013dua}, we assumed that some sort of power corrections such as FSIs amount
to giving a large strong phase $\delta$ to the nonresonant component of  $\la K^-\pi^+|
\bar s d|0\ra$
\be \label{eq:Kpime}
 \la K^-(p_1)\pi^+(p_2)|\bar s d|0\ra^{\rm NR}
 = \frac{v}{3}(3 F_{\rm NR}+2F'_{\rm NR})+\sigma_{_{\rm NR}}
 e^{-\alpha s_{12}}e^{i\delta}.
\en
We found that $\delta\approx \pm\pi$ will enable us to accommodate both branching fractions and \CP asymmetry simultaneously. In practice, we use
\be \label{eq:KpimeNew1}
 \la K^-(p_1)\pi^+(p_2)|\bar s d|0\ra^{\rm NR}
 &\approx& \frac{v}{3}(3 F_{\rm NR}+2F'_{\rm NR})+\sigma_{_{\rm NR}}
 e^{-\alpha s_{12}}e^{i\pi}\left(1+4{m_K^2-m_\pi^2\over s_{12}}\right).
\en
Our calculated nonresonant rate in $B^-\to K^-\pi^+\pi^-$ is consistent with the Belle measurement, but larger than that of BaBar. It
is of the same order
of magnitude as that in $B^-\to K^+K^-K^-$ decays. Indeed, this is what we
will expect.
The reason why the nonresonant fraction is as large as 90\% in
$K\!K\!K$ decays, but becomes only $(17\sim 40)\%$ in $K\pi\pi$
channels (see Table \ref{tab:BRexpt}) can be explained as follows.
Since the $K\!K\!K$ channel receives resonant
contributions only from $\phi$ and $f_{0}$ mesons, while $K^*,
K^*_{0},\rho,f_{0}$ resonances contribute to $K\pi\pi$ modes,
this explains why the nonresonant fraction is of order 90\% in the
former and becomes of order 40\% or smaller in the latter.

Finally, we wish to stress again that the predicted total rate of $B^-\to K^-\pi^+\pi^-$ is smaller than the measurements of both BaBar and Belle. This is ascribed to the fact that the calculated $K_0^*(1430)\pi^-$ in naive factorization is too small by a factor of 3.

\begin{table}[!]
\footnotesize{
\caption{Branching fractions (in units of $10^{-6}$) of resonant and
nonresonant (NR) contributions to $B^-\to \ov K^0\pi^-\pi^0$, $B^-\to K^-\pi^0\pi^0$, $\ov B^0\to\ov K^0\pi^+\pi^-$ and $\ov B^0\to K^-\pi^+\pi^0$.  Note that  the BaBar result for $K_0^{*-}(1430)\pi^+$  in \cite{BaBarK0pippim},
all the BaBar results in \cite{BaBarKppimpi0} and Belle results  in \cite{BelleKppimpi0} are their
absolute ones. We have converted them into the product branching
fractions, namely, $\B(B\to Rh)\times \B(R\to hh)$. }
\begin{ruledtabular}
\begin{tabular}{l l l l} \label{tab:otherKpipi}
$B^-\to \ov K^0\pi^-\pi^0$
\\
Decay mode~~
  & BaBar \cite{BaBar:Kstpi}
  &
  & Theory
\\
\hline
$K^{*-}\pi^0$ & $6.1\pm0.9\pm0.4^{+0.2}_{-0.3}$ &
  & $4.7^{+0.0+1.0+0.1}_{-0.0-0.9-0.1}$ \\
$\ov K^{*0}\pi^-$ & $4.9\pm0.9\pm0.4^{+0.2}_{-0.3}$ &
  & $4.1^{+0.0+1.0+0.0}_{-0.0-0.9-0.0}$
  \\
$K^{*-}_0(1430)\pi^0$ & $10.7\pm1.5\pm0.9^{+0.0}_{-1.1}$ &
  & $5.6^{+0.0+1.6+0.0}_{-0.0-1.4-0.0}$ \\
$\ov K^{*0}_0(1430)\pi^-$ & $15.5\pm1.5\pm1.9^{+0.8}_{-0.8}$ &
  & $5.4^{+0.0+1.7+0.0}_{-0.0-1.4-0.0}$
  \\
$\rho^-\ov K^0$ & $9.4\pm1.6\pm1.1^{+0.0}_{-2.6}$ &
  & $5.9^{+0.0+2.5+0.0}_{-0.0-0.9-0.0}$
  \\
NR & &
  & $9.5^{+0.3+6.3+0.0}_{-0.3-3.6-0.0}$
  \\
\hline
Total & $45.0\pm2.6\pm3.0^{+8.6}_{-0.0}$ &
  & $28.5^{+0.2+12.1+0.0}_{-0.3-~7.4-0.0}$
  \\
\hline \hline
$ B^-\to K^-\pi^0\pi^0$
  \\
Decay mode~~
  & BaBar \cite{BaBarKmpi0pi0}
  &
  & Theory
  \\
\hline
$K^{*-}\pi^0$
  & $2.7\pm0.5\pm0.4$
  &
  &  $2.5^{+0.0+0.6+0.0}_{-0.0-0.5-0.0}$
  \\
$K^{*-}_0(1430)\pi^0$
  & $$
  &
  & $2.4^{+0.0+0.8+0.0}_{-0.0-0.7-0.0}$%%
  \\
$f_0(980)K^-$
  & $2.8\pm0.6\pm0.5$
  &
  & $3.3^{+0.0+0.8+0.0}_{-0.0-0.6-0.0}$ %%
  \\
NR
  &
  &
  & $5.9^{+0.0+2.6+0.0}_{-0.0-1.9-0.0}$ %%
  \\
\hline
Total
  & $16.2\pm1.2\pm1.5$
  &
  &  $13.3^{+0.1+4.6+0.0}_{-0.0-3.5-0.0}$ %%
  \\
\hline\hline
$\ov B^0\to \ov K^0\pi^+\pi^-$
  \\
Decay mode~~
  & BaBar \cite{BaBarK0pippim}
  &  Belle \cite{BelleK0pipi}
  & Theory
  \\
\hline
$K^{*-}\pi^+$
  & $5.52^{+0.61}_{-0.54}\pm0.35\pm0.41$
  & $5.6\pm0.7\pm0.5^{+0.4}_{-0.3}$
  &  $6.8^{+0.0+1.7+0.1}_{-0.0-1.5-0.1}$ %%
  \\
$K^{*-}_0(1430)\pi^+$
  & $18.5^{+1.4}_{-1.1}\pm1.0\pm0.4\pm2.0$
  & $30.8\pm2.4\pm2.4^{+0.8}_{-3.0}$
  & $10.6^{+0.0+3.0+0.0}_{-0.0-2.6-0.0}$ %%
  \\
$\rho^0\ov K^0$
  & $4.37^{+0.70}_{-0.61}\pm0.29\pm0.12$
  & $6.1\pm1.0\pm0.5^{+1.0}_{-1.1}$
  & $3.9^{+0.0+1.9+0.0}_{-0.0-0.9-0.0}$ %%
  \\
$f_0(980)\ov K^0$
  & $6.92\pm0.77\pm0.46\pm0.32$
  & $7.6\pm1.7\pm0.7^{+0.5}_{-0.7}$
  & $6.0^{+0.0+1.5+0.0}_{-0.0-1.2-0.0}$%%
  \\
$f_2(1270)\ov K^0$
  & $1.15^{+0.42}_{-0.35}\pm0.11\pm0.35$
  &
  \\
NR
  & $11.1^{+2.5}_{-1.0}\pm0.9$
  & $19.9\pm2.5\pm1.6^{+0.7}_{-1.2}$
  & $15.2^{+0.2+7.9+0.0}_{-0.2-5.2-0.0}$ %%
  \\
\hline
Total
  & $50.2\pm1.5\pm1.8$
  & $47.5\pm2.4\pm3.7$
  & $40.0^{+0.1+16.9+0.1}_{-0.1-11.2-0.1}$ %%
  \\
\hline\hline
$\ov B^0\to K^-\pi^+\pi^0$
  \\
Decay mode~~
  & BaBar \cite{BaBarKppimpi0}
  & Belle \cite{BelleKppimpi0}
  & Theory
  \\
\hline
$K^{*-}\pi^+$
  & $2.7\pm0.4\pm0.3$
  & $4.9^{+1.5+0.5+0.8}_{-1.5-0.3-0.3}$
  & $3.5^{+0.0+0.9+0.1}_{-0.0-0.8-0.1}$ %%
  \\
$\ov K^{*0}\pi^0$
  & $2.2\pm0.3\pm0.3$
  & $<2.3$
  & $3.0^{+0.0+0.9+0.0}_{-0.0-0.8-0.0}$ %%
  \\
$K^{*-}_0(1430)\pi^+$
  & $8.6\pm0.8\pm1.0$
  &
  & $5.1^{+0.0+1.5+0.0}_{-0.0-1.3-0.0}$ %%
  \\
$\ov K^{*0}_0(1430)\pi^0$
  & $4.3\pm0.3\pm0.7$
  &  &
  $4.2^{+0.0+1.4+0.0}_{-0.0-1.2-0.0}$ %%
  \\
$\rho^+K^-$
  & $6.6\pm0.5\pm0.8$
  & $15.1^{+3.4+1.4+2.0}_{-3.3-1.5-2.1}$
  & $6.5^{+0.0+2.7+0.1}_{-0.0-1.1-0.1}$ %%
  \\
NR
  & $7.6\pm0.5\pm1.0$
  & $5.7^{+2.7+0.5}_{-2.5-0.4}<9.4$
  & $9.2^{+0.3+5.9+0.0}_{-0.4-3.4-0.0}$ %%
\\ \hline
Total
  & $38.5\pm1.0\pm3.9$
  & $36.6^{+4.2}_{-4.1}\pm3.0$
  & $26.6^{+0.3+13.3+0.1}_{-0.4-~7.8-0.1}$ %%
  \\
\end{tabular}
\end{ruledtabular}
 }
\end{table}

\subsubsection{$B^-\to K^+K^-\pi^-$}
Applying $U$-spin symmetry to Eq. (\ref{eq:KpimeNew1}) leads to
\be \label{eq:KpimeNew2}
 \la K^+(p_1)\pi^-(p_2)|\bar d s|0\ra^{\rm NR}
 &\approx & \frac{v}{3}(3 F_{\rm NR}+2F'_{\rm NR})+\sigma_{_{\rm NR}}
 e^{-\alpha s_{12}}e^{i\pi}\left(1-4{m_K^2-m_\pi^2\over s_{12}}\right),
\en
which will be used to describe $B\to K\ov K\pi$ decays.
Contrary to naive expectation, $s\bar s$ resonant contributions to the tree-dominated $B^-\to K^+K^-\pi^-$ decay are strongly suppressed. The only relevant factorizable amplitude which involves the $s\bar s$ current is given by (see Eq. (5.1) of \cite{Cheng:2013dua})
\be \label{eq:AKpKmpim}
\la \pi^-|(\bar d b)_{V-A}|B ^-\ra
                   \la K^+ K^-|(\bar s s)_{V-A}|0\ra
    \bigg[a_3+a_5-\frac{1}{2}(a_7+a_9)\bigg].
\en
The smallness of the penguin coefficients $a_{3,5,7,9}$ indicates negligible $s\bar s$ resonant contributions.
Indeed, no clear $\phi(1020)$ signature is observed in the mass region $m_{K^+K^-}^2$ around 1 GeV$^2$
\cite{LHCb:pippippim}. The branching fraction of  the  two-body decay $B^-\to\phi\pi^-$
is expected to be very small, of order $4.3\times 10^{-8}$. It is induced mainly from $B^-\to\omega\pi^-$ followed by a
small $\omega-\phi$ mixing \cite{CC:Bud}.

The predicted nonresonant fraction is very sizable about 55\% in $B^-\to K^+K^-\pi^-$ even it is a tree-dominated mode. This should be checked experimentally.

\subsubsection{$B^-\to \pi^+\pi^-\pi^-$}

The current-induced nonresonant contributions to  the tree-dominated $B^-\to \pi^+\pi^-\pi^-$ decay are suppressed by the smallness of the penguin Wilson coefficients $a_6$ and $a_8$. Therefore, the nonresonant component of this decay is predominated by the transition process, and its measurement provides an ideal place to constrain the parameter $\alpha_{_{\rm NR}}$.

\subsubsection{Other $B\to K\pi\pi$ decays}

Branching fractions of resonant and
nonresonant (NR) contributions to other $B\to K\pi\pi$ decays such as $B^-\to \ov K^0\pi^-\pi^0$, $B^-\to K^-\pi^0\pi^0$, $\ov B^0\to\ov K^0\pi^+\pi^-$ and $\ov B^0\to K^-\pi^+\pi^0$ are shown in Table \ref{tab:otherKpipi}.
Except the first channel, the other three have been studied before in \cite{Cheng:2013dua}.
In order to improve the discrepancy between theory and experiment for penguin-dominated $V\!P$ modes in \cite{Cheng:2013dua}, we shall introduce penguin annihilation given in Eq. (\ref{eq:beta}). In general, the predicted $K^*\pi$ and $\rho K$ rates are now consistent with experiment. However, the calculated $K_0^*(1430)\pi$ rates are still too small. This explains why the calculated total branching fractions are smaller than experiment, especially for $B^-\to \ov K^0\pi^-\pi^0$ due to the presence of two $K_0^*(1430)\pi$ modes.

In \cite{Cheng:2013dua} we have made predictions for the resonant and nonresonant contributions to $\ov B^0\to\pi^+\pi^-\pi^0, \ov
K^0\pi^0\pi^0, K_SK^\pm\pi^\mp$.
The $\pi^+\pi^-\pi^0$ mode is predicted to have a rate
larger than $\pi^+\pi^-\pi^-$ even though the former involves a
$\pi^0$  and has no identical particles in the final state. This is because while the latter is
dominated by the $\rho^0$ pole, the former receives
$\rho^\pm$ and $\rho^0$ resonant contributions.

\section{Direct \CP asymmetries}

Experimental measurements of inclusive and regional direct \CP violation by LHCb for various charmless three-body $B$ decays are collected in Table
\ref{tab:CPdata}.
\CP asymmetries of the pair $\pi^-\pi^+\pi^-$ and $K^-K^+K^-$ are of opposite signs, and likewise for the pair $K^-\pi^+
\pi^-$ and $\pi^-K^+K^-$. This can be understood in terms of U-spin symmetry,
which leads to the relation \cite{Bhattacharya,XGHe}
\be
R_1\equiv {\A_{C\!P}(B^-\to \pi^-\pi^+\pi^-)\over  \A_{C\!P}(B^-\to K^-K^+K^-)} &=& -{\Gamma(B^-\to K^-K^+K^-)
\over \Gamma(B^-\to\pi^-\pi^+\pi^-)},
\en
and
\be
R_2\equiv {\A_{C\!P}(B^-\to \pi^-K^+K^-)\over \A_{CP}(B^-\to K^-\pi^+\pi^-)} &=& -{\Gamma(B^-\to K^-\pi^+\pi^-)
\over \Gamma(B^-\to\pi^-K^+K^-)}.
\en
The predicted signs of the ratios $R_1$ and $R_2$ are confirmed by experiment. However,  because of the momentum dependence of 3-body decay amplitudes,  U-spin or flavor SU(3) symmetry does not lead to any testable relations between $\A_{C\!P}(\pi^- K^+ K^-)$ and $\A_{C\!P}
(\pi^-\pi^+\pi^-)$ and between $\A_{C\!P}(K^- \pi^+ \pi^-)$ and $\A_{C\!P}(K^+K^-K^-)$. That is, symmetry argument alone does not give hints at the relative sign of \CP asymmetries in the pair of $\Delta S=0 (1)$ decay.

The LHCb data in Table \ref{tab:CPdata} indicate that decays involving a $K^+K^-$ pair have a larger \CP asymmetry ($\A_{C\!P}^{\rm incl}$ or $\A_{C\!P}^{\rm resc}$)
than their partner channels. The asymmetries are positive for channels with a $\pi^+\pi^-$ pair and negative for those with a $K^+K^-$ pair.
In other words, when $K^+K^-$ is replaced by $\pi^+\pi^-$, \CP asymmetry is flipped in sign. This observation appears to imply that final-state rescattering may play an important role for direct \CP violation. It has been conjectured that maybe the final rescattering between $\pi^+\pi^-$ and $K^+K^-$ in conjunction with {\it CPT} invariance is responsible for the sign change \cite{Bhattacharya,Bigi,Bediaga}.
However, the implication of the {\it CPT} theorem for \CP asymmetries at the hadron level in exclusive or semi-inclusive reactions is more complicated
and remains mostly unclear \cite{AtwoodCPT}.

It is well known that one needs nontrivial strong and weak phase differences to produce  partial rate \CP
asymmetries. In this work, the strong phases arise from the
effective Wilson coefficients $a_i^p$ listed in Eq. (2.3) of \cite{Cheng:2013dua}, the Breit-Wigner expression for resonances and the penguin
matrix elements of scalar densities.  It has been established that the strong phase in the penguin coefficients $a_6^p$ and $a_8^p$ comes from the Bander-Silverman-Soni mechanism \cite{BSS}. There are two sources for the phase in the penguin matrix elements of scalar densities: $\sigma_{\rm NR}$ and $\delta$ for $K\pi$--vacuum matrix elements.

In the literature, most of the theory studies concentrate on the resonant effects on \CP violation. For example, the authors of \cite{Zhang,Bhattacharya}  considered the
possibility of having a large local \CP violation in $B^-\to \pi^+\pi^-\pi^-$ resulting from the interference of the resonances
$f_0(500)$ and  $\rho^0(770)$. A similar mechanism has been applied to the decay $B^-\to K^-\pi^+\pi^-$ \cite{Zhang}.

In this work, we shall take into account both resonant and nonresonant amplitudes
simultaneously and work out their contributions and interference to branching fractions and \CP violation in details.

\subsection{$CP$ asymmetries due to resonant and nonresonant contributions}

%%%%%%%%%%%%%%%%%%%%%%%%%%%%%%%%%%%
\begin{table}[t]
\caption{Predicted inclusive and regional \CP asymmetries (in \%) for various charmless three-body $B$ decays. Two local
regions of interest for regional \CP asymmetries are the low-mass regions specified in Eq. (\ref{eq:KKKlocalCP}) for $\A_{C\!P}^{\rm
incl}$ and the rescattering region of $m_{\pi\pi}$ and $m_{K\!\ov K}$ between 1.0 and 1.5 GeV for $\A_{C\!P}^{\rm
resc}$. Resonant (RES) and nonresonant (NR) contributions to direct \CP asymmetries are considered.}
\begin{ruledtabular} \label{tab:CPasy}
\begin{tabular}{l  c c c c }
  & $\pi^-\pi^+\pi^-$
  & $K^+K^-\pi^-$
  & $K^-\pi^+\pi^-$
  & $K^+K^-K^-$
  \\ \hline
$(\A_{C\!P}^{\rm incl})_{\rm NR}$
  & $25.0^{+4.4+2.1+0.0}_{-2.7-3.1-0.1}$
  & $-25.6^{+2.2+1.7+0.2}_{-3.0-1.1-0.1}$
  & $9.1^{+1.3+2.2+0.1}_{-1.8-2.0-0.1}$
  & $-7.8^{+1.4+1.3+0.1}_{-0.9-1.5-0.1}$ \\
$(\A_{C\!P}^{\rm incl})_{\rm RES}$
  & $5.3^{+0.0+1.6+0.0}_{-0.0-1.3-0.0}$
  & $-16.3^{+0.0+0.9+0.1}_{-0.0-0.8-0.1}$
  & $6.9^{+0.0+2.1+0.1}_{-0.0-1.8-0.1}$
  & $1.2^{+0.0+0.0+0.0}_{-0.0-0.0-0.0}$ \\
$(\A_{C\!P}^{\rm incl})_{\rm NR+RES}$
  & $8.3^{+0.5+1.6+0.0}_{-1.1-1.5-0.0}$
  & $-10.2^{+1.6+1.5+0.1}_{-2.5-1.4-0.1}$
  & $7.3^{+0.2+2.1+0.1}_{-0.2-2.0-0.1}$
  & $-6.0^{+1.8+0.8+0.1}_{-1.2-0.9-0.1}$ \\
$(\A_{C\!P}^{\rm incl})_{\rm expt}$
  & $5.8\pm2.4$
  & $-12.3\pm2.2$
  & $2.5\pm0.9$
  & $-3.6\pm0.8$ \\
  \hline
$(\A_{C\!P}^{\rm low})_{\rm NR}$
  & $58.3^{+3.6+2.6+0.8}_{-3.7-4.0-0.8}$
  & $-25.0^{+2.8+2.7+0.3}_{-5.4-2.5-0.3}$
  & $48.9^{+~7.0+7.6+0.3}_{-10.5-8.2-0.3}$
  & $-13.0^{+2.0+2.8+0.2}_{-1.2-3.2-0.2}$ \\
$(\A_{C\!P}^{\rm low})_{\rm RES}$
  & $4.5^{+0.0+1.6+0.0}_{-0.0-1.2-0.0}$
  & $-4.9^{+0.0+0.5+0.0}_{-0.0-0.4-0.0}$
  & $57.1^{+0.0+~7.9+0.9}_{-0.0-16.6-0.9}$
  & $1.6^{+0.0+0.1+0.0}_{-0.0-0.1-0.0}$ \\
$(\A_{C\!P}^{\rm low})_{\rm NR+RES}$
  & $21.9^{+0.5+3.0+0.0}_{-0.4-3.3-0.1}$
  & $-17.5^{+0.6+1.7+0.1}_{-0.9-1.5-0.1}$
  & $49.4^{+0.7+~9.4+0.8}_{-1.0-14.2-0.8}$
  & $-16.8^{+3.5+2.8+0.2}_{-2.3-3.2-0.2}$ \\
$(\A_{C\!P}^{\rm low})_{\rm expt}$
  & $58.4\pm9.7$
  & $-64.8\pm7.2$
  & $67.8\pm8.5$
  & $-22.6\pm2.2$ \\
  \hline
$(\A_{C\!P}^{\rm resc})_{\rm NR}$
  & $36.7^{+6.2+3.2+0.1}_{-3.7-4.6-0.2}$
  & $-27.7^{+3.1+3.0+0.4}_{-5.9-2.7-0.4}$
  & $31.8^{+4.6+4.6+0.3}_{-6.7-4.5-0.3}$
  & $-10.8^{+1.8+2.2+0.2}_{-1.2-2.5-0.2}$ \\
$(\A_{C\!P}^{\rm resc})_{\rm RES}$
  & $7.0^{+0.0+1.8+0.0}_{-0.0-1.5-0.0}$
  & $-5.6^{+0.0+0.5+0.0}_{-0.0-0.4-0.0}$
  & $1.1^{+0.0+0.6+0.0}_{-0.0-0.5-0.0}$
  & $0.96^{+0.00+0.02+0.01}_{-0.00-0.02-0.01}$\\
$(\A_{C\!P}^{\rm resc})_{\rm NR+RES}$
  & $13.4^{+0.5+2.0+0.0}_{-1.1-2.1-0.0}$
  & $-20.4^{+1.2+2.0+0.2}_{-1.8-1.8-0.2}$
  & $4.1^{+0.2+0.9+0.0}_{-0.3-0.9-0.0}$
  & $-3.8^{+1.5+0.5+0.1}_{-1.0-0.5-0.1}$ \\
$(\A_{C\!P}^{\rm resc})_{\rm expt}$
  & $17.2\pm2.7$
  & $-32.8\pm4.1$
  & $12.1\pm2.2$
  & $-21.1\pm1.4$ \\
\end{tabular}
\end{ruledtabular}
\end{table}
%%%%%%%%%%%%%%%%%%%%%%%%%%%%%%%%

Following the framework of \cite{CCS:nonres,Cheng:2013dua} we present in Table
\ref{tab:CPasy} the calculated results of inclusive and regional \CP asymmetries in our model.  We consider both resonant and nonresonant mechanisms and their interference. For nonresonant contributions, direct \CP violation arises solely from the interference of tree and
penguin nonresonant amplitudes. For example, in the absence of resonances, \CP asymmetry in $B^-\to K^-\pi^+\pi^-$
stems mainly from the interference of the nonresonant tree amplitude $\la \pi^+\pi^-|(\bar u b)_{V-A}|B^-\ra^{\rm NR}\la K^-|(\bar
su)_{V-A}|0\ra$ and the nonresonant penguin amplitude $\la \pi^-|\bar db|B^-\ra\la K^-\pi^+|\bar sd|0\ra^{\rm NR}$.

It is clear from Table \ref{tab:CPasy} that  nonresonant \CP violation is usually much larger than the resonant one and that the interference effect is generally quite significant. If nonresonant contributions are turned off in the $K^+K^-K^-$ mode, the predicted asymmetries will be wrong in sign when compared with experiment. This is not a surprise because $B^-\to K^+K^-K^-$ is predominated by the nonresonant background. The magnitude and the sign of its \CP asymmetry should be governed by the nonresonant term.

Large local \CP asymmetries $\A_{C\!P}^{\rm low}$ in three-body charged $B$ decays have been observed by LHCb in the low mass regions specified in Eq. (\ref{eq:KKKlocalCP}). If
intermediate resonant states are not associated in these low-mass regions, it is natural to expect that the Dalitz plot is
governed by nonresonant contributions.  It is evident from Table \ref{tab:CPasy} that except the mode $K^+K^-\pi^-$, \CP violation in the low mass region is indeed dominated by the nonresonant background. In our model we find large nonresonant contributions to \CP asymmetries for $B^-\to \pi^+\pi^-\pi^-, \pi^+\pi^-K^-$, of order $0.58$ and $0.49$, respectively. Likewise, large $(\A_{C\!P}^{\rm low})_{\rm NR}=(51.9^{+1.08+0.27}_{-0.91-0.32})\%$ for the former mode was also obtained in the pQCD approach \cite{Wang:2014ira}.

From Table \ref{tab:CPasy}, it is evident that except the $K^+K^-K^-$ mode, the resonant contributions to integrated inclusive \CP asymmetries are of the same sign and similar magnitudes as $\A_{C\!P}^{\rm incl}$. For $\pi^+\pi^-\pi^-$, resonant \CP violation is dominated by the $\rho^0$, $\A_{C\!P}(\rho^0\pi^-)=0.059^{+0.012}_{-0.010}$, which is close to the resonance-induced integrated asymmetry $(\A_{C\!P}^{\rm incl})_{\rm RES}=(5.3^{+1.6}_{-1.3})\%$. However, there is an issue about the theoretical predictions of $\A_{C\!P}(\rho^0\pi^-)$, which will be addressed in detail below.
The resonant \CP asymmetry in $B^-\to K^-\pi^+\pi^-$ is governed by the $\rho^0$ with $\A_{C\!P}(\rho^0 K^-)=0.65^{+0.10}_{-0.21}$, while the world average of measurements is $0.37\pm0.11$ \cite{HFAG}. For $K^+K^-\pi^-$, we have the dominant contributions from $\A_{C\!P}(K^{*0} K^-)=-28.4\%$ and $\A_{C\!P}(K^{*0}_0(1430) K^-)=-19.2\%$. For $K^+K^-K^-$, the main contributions to $(\A_{C\!P}^{\rm incl})_{\rm RES}$ arise from $\phi K^-,f_0(1500)K^-,f_0(1710)K^-$, all give positive contributions.  The observed negative  $\A_{C\!P}^{\rm incl}(K^+K^-K^-)$ is a strong indication of the importance of nonresonant effects. This is reinforced by the fact that the predicted $(\A_{C\!P}^{\rm low})_{\rm RES}$ and  $(\A_{C\!P}^{\rm resc})_{\rm RES}$ by resonances alone are usually too small compared to the data, especially for the former.

\subsection{Discussions}
Although our model based on factorization describes the observed asymmetries reasonably well, in the following we would like to address several related issues.

%%%%%%%%%%%%%%%%%%%%%%%%%%%%%%%%%%%
\begin{table}[t]
\caption{Same as Table \ref{tab:CPasy} except that the strong phase $\delta$ defined in Eq. (\ref{eq:Kpime}) for $K\pi$ matrix element of scalar density is set to zero. The decays $B^-\to \pi^+\pi^-\pi^-$ and $K^+K^-K^-$ are not affected by the phase $\delta$ .}
\begin{ruledtabular} \label{tab:CPdelta}
\begin{tabular}{l  c c }
  & $K^+K^-\pi^-$
  & $K^-\pi^+\pi^-$
  \\ \hline
$(\A_{C\!P}^{\rm incl})_{\rm NR}$
  & $17.4^{+0.7+1.7+0.0}_{-1.0-2.9-0.1}$
  & $-3.5^{+0.8+1.1+0.1}_{-0.6-1.3-0.0}$
  \\
$(\A_{C\!P}^{\rm incl})_{\rm RES}$
  & $-16.3^{+0.0+0.9+0.1}_{-0.0-0.8-0.1}$
  & $6.9^{+0.0+2.1+0.1}_{-0.0-1.8-0.1}$
 \\
$(\A_{C\!P}^{\rm incl})_{\rm NR+RES}$
  & $4.9^{+0.7+0.9+0.1}_{-0.8-0.6-0.1}$
  & $-0.8^{+0.7+0.6+0.0}_{-0.5-0.3-0.0}$
 \\
$(\A_{C\!P}^{\rm incl})_{\rm expt}$
  & $-12.3\pm2.2$
  & $2.5\pm0.9$
  \\
  \hline
$(\A_{C\!P}^{\rm low})_{\rm NR}$
  & $22.3^{+5.3+2.6+0.0}_{-2.8-2.9-0.1}$
  & $-19.0^{+1.5+5.0+0.4}_{-0.7-5.9-0.3}$
  \\
$(\A_{C\!P}^{\rm low})_{\rm RES}$
  & $-4.9^{+0.0+0.5+0.0}_{-0.0-0.4-0.0}$
  & $57.1^{+0.0+~7.9+0.9}_{-0.0-16.6-0.9}$
 \\
$(\A_{C\!P}^{\rm low})_{\rm NR+RES}$
  & $4.6^{+0.7+0.6+0.0}_{-0.4-0.8-0.0}$
  & $40.7^{+3.2+5.0+0.3}_{-2.4-8.6-0.4}$
  \\
$(\A_{C\!P}^{\rm low})_{\rm expt}$
  & $-64.8\pm7.2$
  & $67.8\pm8.5$
  \\
  \hline
$(\A_{C\!P}^{\rm resc})_{\rm NR}$
  & $25.2^{+5.9+2.8+0.0}_{-3.1-3.2-0.1}$
  & $-11.5^{+1.6+3.2+0.2}_{-0.9-3.8-0.2}$
 \\
$(\A_{C\!P}^{\rm resc})_{\rm RES}$
  & $-5.6^{+0.0+0.5+0.0}_{-0.0-0.4-0.0}$
  & $1.1^{+0.0+0.6+0.0}_{-0.0-0.5-0.0}$
  \\
$(\A_{C\!P}^{\rm resc})_{\rm NR+RES}$
  & $10.1^{+1.2+1.3+0.0}_{-0.7-1.5-0.1}$
  & $-6.4^{+1.0+0.3+0.1}_{-0.7-0.1-0.1}$
  \\
$(\A_{C\!P}^{\rm resc})_{\rm expt}$
  & $-32.8\pm4.1$
  & $12.1\pm2.2$
  \\
\end{tabular}
\end{ruledtabular}
\end{table}
%%%%%%%%%%%%%%%%%%%%%%%%%%%%%%%%

\subsubsection{\CP asymmetry induced by interference}
\CP asymmetry of the $B^-\to\pi^+\pi^-\pi^-$ decay in the low-mass region of $m(\pi^+\pi^-)_{\rm low}$ is observed to change sign at a value of $m(\pi^+\pi^-)_{\rm low}$ close to the $\rho(770)$ resonance. This change of sign occurs for both $\cos\theta>0$ and $\cos\theta<0$ (see Fig. 4 of \cite{LHCb:2014}), where $\theta$ is the angle between the momenta of the unpaired hadron and the resonance decay product with the same-sign charge. Likewise, the Dalitz \CP asymmetry of $B^-\to K^-\pi^+\pi^-$ has two zeros in the $m(\pi^+\pi^-)$ distribution. In the $\cos\theta<0$ region there is a zero around the $\rho(770)$ mass and another one around the $f_0(980)$ meson mass  (see Fig. 5 of \cite{LHCb:2014}). However, in the region of $\cos\theta>0$, a clear change of sign is only seen around the $f_0(980)$ mass.

In this work we do see the sign change of \CP asymmetry in the decay $B^-\to\pi^+\pi^-\pi^-$ for $\cos\theta<0$ but not for $\cos\theta>0$. The former arises from the interference of $\rho(770)$ with the nonresonant background.  The sign change is ascribed to the real part of the Breit-Wigner propagator of the $\rho(770)$ which reads
\be
{s-m_\rho^2\over (s-m_\rho^2)^2+m_\rho^2\Gamma_\rho^2(s)}.
\en
It is not clear to us why we did not see the zero for $\cos\theta>0$. As for $B^-\to K^-\pi^+\pi^-$, the interference between $\rho(770)$ and $f_0(980)$ has a real component proportional to
\be
{(s-m_\rho^2)(s-m_{f_0}^2)\over [(s-m_\rho^2)^2+m_\rho^2\Gamma_\rho^2(s)][(s-m_{f_0}^2)^2+m_{f_0}^2\Gamma_{f_0}^2(s)]}.
\en
This gives to two zeros: one at $s=m_{\rho(770)}^2$ and the other at $s=m_{f_0(980)}^2$. However, we only see a sign change around $f_0(980)$ but not $\rho(770)$ for $\cos\theta<0$ and do not see any zero for $\cos\theta>0$. It is possible that the zeros are contaminated or washed out by other contributions. We are going to investigate this issue.

\subsubsection{Strong phase $\delta$}

We now discuss in more details why we need to introduce an additional phase $\delta$ to the matrix element of scalar density $\la K^-\pi^+|\bar sd|0\ra$ given in Eq. (\ref{eq:Kpime}). First, we notice that
the calculated integrated \CP asymmetries $(8.3^{+1.7}_{-1.9})\%$ for $\pi^+\pi^-\pi^-$ and $(-6.0^{+2.0}_{-1.5})\%$
for $K^+K^-K^-$ (see Table \ref{tab:CPasy}) are consistent with LHC measurements  in both sign and magnitude.
\footnote{Before the LHCb measurements of \CP violation in three-body $B$ decays, the predicted \CP asymmetries in various charmless three-body $B$ decays can be found in Table XVII of \cite{CCS:nonres}.}
As discussed in passing and in \cite{Cheng:2013dua},  when
the unknown two-body matrix elements of scalar densities $\la K\pi|\bar s q|0\ra$ and $\la \pi K|\bar s q|0\ra$ are related to $\la K\bar K|\bar s s|0\ra$ via SU(3) symmetry so that $\la K^-\pi^+|\bar sd|0\ra=\la K^+\pi^-|\bar ds|0\ra=\la
K^+K^-|\bar ss|0\ra$, the calculated nonresonant and total rates of $B^-\to K^-\pi^+\pi^-$ will be too large compared to experiment [see the discussions after Eq. (\ref{eq:Kpim.e.SU3})]. Moreover,
the predicted \CP violation $\A_{C\!P}^{\rm incl}(K^-\pi^+\pi^-)=(-0.8^{+0.9}_{-0.6})\%$ and $\A_{C\!P}^{\rm incl}(K^+K^-\pi^-)=(4.9^{+1.1}_{-1.0})\%$ are wrong in
sign when confronted with experiment. Since the partial rate asymmetry arises from the interference between tree and penguin amplitudes and since nonresonant penguin
contributions to the penguin-dominated decay $K^-\pi^+\pi^-$ are governed by the matrix element $\la K^-\pi^+|\bar sd|
0\ra$, it is thus conceivable that a strong phase $\delta$ in  $\la K^-\pi^+|\bar sd|0\ra$ induced from some sort of power corrections might flip the sign of \CP asymmetry.

It is clear from Table \ref{tab:CPdelta} that the reason why the predicted inclusive and regional \CP asymmetries (except $\A_{C\!P}^{\rm low}(K^-\pi^+\pi^-)$) all are erroneous in sign when $\delta$ is set to zero is ascribed to the nonresonant contributions which are opposite in sign to the experimental measurements. By comparing Tables \ref{tab:CPdelta} and \ref{tab:CPasy}, we see that when $\delta$ is set to $\approx \pm\pi$ preferred by the data, \CP asymmetries induced from nonresonant components will flip the sign as $e^{\pm i\pi}=-1$. Consequently, this in turn will lead to the correct sign for the predicted asymmetries. As stressed in \cite{Cheng:2013dua},
we have implicitly assumed that power corrections will not affect \CP violation in $\pi^+\pi^-\pi^-$ and $K^+K^-K^-$.

Finally we would like to remark that unlike the global weak phases, strong phases such as $\delta$ and the Breit-Wigner phase are local ones, namely they are energy and channel dependent. For example, when we study {\it CP}-asymmetry Dalitz distributions in some large invariant mass regions (see subsection III.4 below), we find that $\delta$ needs to vanish in the large invariant mass region for $B^-\to K^+K^-\pi^-$ in order to accommodate the observation.

\subsubsection{Final-state rescattering}
As shown in Table \ref{tab:CPdelta}, the calculated integrated and local \CP asymmetries  $\A_{C\!P}^{\rm incl}$, $\A_{C\!P}^{\rm low}$ and $\A_{C\!P}^{\rm resc}$ for $B^-\to K^+K^-\pi^-, K^-\pi^+\pi^-$ with $\delta=0$ are wrong in
sign when confronted with experiment. Since direct \CP
violation in charmless two-body $B$ decays can be significantly
affected by final-state rescattering \cite{CCSfsi}, it is natural
to hope that final-state rescattering effects in three-body $B$ decays may resolve the discrepancy.  For example, the sign of the \CP asymmetry in the two-body decay $\bar B^0\to K^-\pi^+$ can be flipped by the presence of long-distance rescattering of charming penguins \cite{CCSfsi}.

Just as the example of $\ov B^0\to K^-\pi^+$ whose \CP violation is originally predicted to have wrong sign in naive factorization and gets a correct sign after power corrections such as final-state interactions or penguin annihilation, are taken into account, it will be very interesting to see  an explicit demonstration of the sign flip of $\A_{CP}(K^-\pi^+\pi^-)$ and $\A_{CP}(\pi^-K^+K^-)$ when the final-state rescattering of $\pi\pi\leftrightarrow K\ov K$ is turned on.

Here we shall follow the work of \cite{Pelaez:2004vs} (also the same framework adapted in \cite{Bediaga:2015}) to describe the inelastic $\pi\pi\leftrightarrow K\bar K$ rescattering process and consider this final-state rescattering effect on inclusive and local \CP violation.

%%%%%%%%%%%%%%%%%%%%%%%%%%%%%%%%%%%
\begin{table}[t]
\caption{Predicted inclusive and regional \CP asymmetries (in \%) for various charmless three-body $B$ decays in the presence of $\pi^+\pi^-\leftrightarrow K^+K^-$ final-state rescattering. We have set $\delta$ to zero. Only the central values of the final-state interaction (FSI) effects are quoted here.}
\begin{ruledtabular} \label{tab:Afsi}
\begin{tabular}{l  c c c c }
  & $\pi^-\pi^+\pi^-$
  & $K^+K^-\pi^-$
  & $K^-\pi^+\pi^-$
  & $K^+K^-K^-$
  \\ \hline
$(\A_{C\!P}^{\rm incl})_{\rm NR+RES}$
  & $8.3^{+0.3+1.6+0.0}_{-1.1-1.5-0.0}$
  & $4.9^{+0.7+0.9+0.1}_{-0.8-0.6-0.1}$
  & $-0.8^{+0.7+0.6+0.0}_{-0.5-0.3-0.0}$
  & $-6.0^{+1.8+0.8+0.1}_{-1.2-0.9-0.1}$ \\
$(\A_{C\!P}^{\rm incl})_{\rm NR+RES+FSI}$
  & $-15.6$
  & $8.1$
  & $0.7$
  & $-6.1$ \\
$(\A_{C\!P}^{\rm incl})_{\rm expt}$
  & $5.8\pm2.4$
  & $-12.3\pm2.2$
  & $2.5\pm0.9$
  & $-3.6\pm0.8$ \\
  \hline
$(\A_{C\!P}^{\rm low})_{\rm NR+RES}$
  & $21.9^{+0.5+3.0+0.0}_{-0.4-3.3-0.1}$
  & $4.6^{+0.7+0.6+0.0}_{-0.4-0.8-0.0}$
  & $40.7^{+3.2+5.0+0.3}_{-2.4-8.6-0.4}$
  & $-16.8^{+3.5+2.8+0.2}_{-2.3-3.2-0.2}$ \\
$(\A_{C\!P}^{\rm low})_{\rm NR+RES+FSI}$
  & $-17.6$
  & $13.2$
  & $2.3$
  & $-16.7$ \\
$(\A_{C\!P}^{\rm low})_{\rm expt}$
  & $58.4\pm9.7$
  & $-64.8\pm7.2$
  & $67.8\pm8.5$
  & $-22.6\pm2.2$ \\
  \hline
$(\A_{C\!P}^{\rm resc})_{\rm NR+RES}$
  & $13.4^{+0.5+2.0+0.0}_{-1.1-2.1-0.0}$
  & $10.1^{+1.2+1.3+0.0}_{-0.7-1.5-0.1}$
  & $-6.4^{+1.0+0.3+0.1}_{-0.7-0.1-0.1}$
  & $-3.8^{+1.5+0.5+0.1}_{-1.0-0.5-0.1}$ \\
$(\A_{C\!P}^{\rm resc})_{\rm NR+RES+FSI}$
  & $10.4$
  & $20.0$
  & $-1.3$
  & $-4.0$ \\
$(\A_{C\!P}^{\rm resc})_{\rm expt}$
  & $17.2\pm2.7$
  & $-32.8\pm4.1$
  & $12.1\pm2.2$
  & $-21.1\pm1.4$ \\
\end{tabular}
\end{ruledtabular}
\end{table}
%%%%%%%%%%%%%%%%%%%%%%%%%%%%%%%%

The general expression of 3-body $B$ decay amplitude under final-state interactions is given by \cite{Chua,Suzuki}
\be
A_i^{\rm FSI}=\sum _{j=1}^n (S^{1/2})_{ij}A_j^{\rm fac}.
\en
We now concentrate on $\pi^+\pi^-$ and $K^+K^-$ final-state rescattering and neglect possible interactions with the third meson under the so-called ``2$+$1" assumption and write
\be
\left(\begin{array}{c} A(B^-\to \pi^+\pi^-P^-) \\ A(B^-\to K^+K^-P^-)
\end{array}\right)^{\rm FSI}=S^{1/2} \left(\begin{array}{c} A(B^-\to \pi^+\pi^-P^-) \\ A(B^-\to K^+K^-P^-)
\end{array}\right)
\en
with $P=\pi,K$. The unitary $S$ matrix reads
\be
S=\left(
\begin{array}{cc}
\eta e^{2i\delta_{\pi\pi}} & i\sqrt{1-\eta^2} e^{i(\delta_{\pi\pi}+\delta_{K\!\bar K})} \\
i\sqrt{1-\eta^2} e^{i(\delta_{\pi\pi}+\delta_{K\!\bar K})} & \eta e^{2i\delta_{K\!\bar K}}
\end{array}
\right),
\en
where the inelasticity parameter $\eta(s)$ is given by \cite{Pelaez:2004vs}
\be
\eta(s)=1-\left(\epsilon_1{k_2\over s^{1/2} }+\epsilon_2{k_2^2\over s}\right)\,{ {M'}^2-s\over s},
\en
with
\be
k_2={\sqrt{s-4m_K^2}\over 2}.
\en
The $\pi\pi$ phase shift has the expression
\be
\delta_{\pi\pi}(s)={1\over 2}\cos^{-1}\left( {\cot^2[\delta_{\pi\pi}(s)]-1\over \cot^2[\delta_{\pi\pi}(s)]+1 }\right),
\en
with
\be
\cot[\delta_{\pi\pi}(s)]=c_0\,{(s-M_s^2)(M_f^2-s)\over M_f^2 s^{1/2} }\,{|k_2|\over k_2^2}.
\en
We shall assume that $\delta_{K\!\bar K}\approx \delta_{\pi\pi}$ in the rescattering region.

To calculate $S^{1/2}$, we note that the $S$-matrix can be recast to the form
\be
S &=& U \left(
\begin{array}{cc}
\eta e^{2i\delta_{\pi\pi}}(\eta-i\sqrt{1-\eta^2}) & 0  \\
0 & \eta e^{2i\delta_{\pi\pi}}(\eta+i\sqrt{1-\eta^2})
\end{array}
\right) U^\dagger \non \\
&=& U e^{2i\delta_{\pi\pi}}\left(
\begin{array}{cc}
e^{-i\phi} & 0  \\
0 & e^{i\phi}
\end{array}
\right) U^\dagger,
\en
with
\be
U={1\over\sqrt{2}}\left(
\begin{array}{cc}
1 & 1  \\
-1 & 1
\end{array}
\right)
\en
and
\be
\phi=\tan^{-1}{\sqrt{1-\eta^2}\over \eta}.
\en
Hence,
\be
S^{1/2}= U e^{i\delta_{\pi\pi}}\left(
\begin{array}{cc}
e^{-i\phi/2} & 0  \\
0 & e^{i\phi/2}
\end{array}
\right) U^\dagger =e^{i\delta_{\pi\pi}} \left(
\begin{array}{cc}
\cos\phi/2 & i\sin\phi/2  \\
i\sin\phi/2 & \cos\phi/2
\end{array}
\right).
\en
Consequently,
\be
A(B^-\to \pi^+\pi^- P^-)^{\rm FSI} &=& e^{i\delta_{\pi\pi}}\Big[\cos(\phi/2)A(B^-\to \pi^+\pi^- P^-)+i\sin(\phi/2)A(B^-
\to K^+K^- P^-)\Big], \non \\
A(B^-\to K^+K^- P^-)^{\rm FSI} &=& e^{i\delta_{\pi\pi}}\Big[\cos(\phi/2)A(B^-\to K^+K^- P^-)+i\sin(\phi/2)A(B^-\to
\pi^+\pi^- P^-)\Big], \non \\
\en
for $P=\pi, K$.

For the numerical results presented in Table \ref{tab:Afsi}, we have used the parameters given in Eqs.~(2.15b') and (2.16) of \cite{Pelaez:2004vs}, namely $M'=1.5$ GeV, $M_s=0.92$ GeV, $M_f=1.32$ GeV, $\epsilon_1=2.4$, $\epsilon_2=-5.5$ and $c_0=1.3$\,.
Unfortunately, our results are rather disappointed: In the presence of the specific final-state rescattering, \CP asymmetries for both $\pi^+\pi^-\pi^-$ and $K^+K^-\pi^-$ are heading to the wrong direction. While $\A_{C\!P}$ is decreased for the former, it is increased for the latter, rendering the discrepancy between theory and experiment even worse. We also see that $\A_{C\!P}(K^+K^-K^-)$ is almost not affected by the rescattering of $\pi\pi$ and $K\bar K$.

Thus far we have confined ourselves to rescattering between $\pi^+\pi^-$ and $K^+K^-$ in $s$-wave configuration.
It is known from two-body $B$ decays that this particular rescattering channel (through annihilation and total annihilation diagrams, see Fig.~1 of \cite{Chua}) cannot be sizeable, or the rescattered $B^0\to K^+K^-$ rate fed from the $B^0\to\pi^+\pi^-$ mode will easily excess the measured rate, which is highly suppressed \cite{PDG}.
In fact, the effect of exchange rescattering is expected to be more prominent \cite{Chua} and one needs to enlarge the rescattering channels.  It is clear that $\pi\pi$ and $KK$ are not confined to the $s$-wave configuration in the three-body decays. Therefore, rescatterings in other partial wave configurations should also be included. Rescatterings between the third meson and other mesons can be relevant. Moreover, other potentially important coupled channels should not be neglected. For example, the decay $B^-\to\pi^+\pi^-\pi^-$ can be produced through the weak decay $B\to D\bar D^*\pi$ followed by the rescattering of $D\bar D^*\pi\to \pi\pi\pi$ and likewise for other three-body decays of $B$ mesons. The intermediate $D^{(*)}_{(s)}\bar D^{(*)}_{(s)}P$ states have large CKM matrix elements  and hence can make significant contributions to \CP violation when coupled to three light pseudoscalar states.

A comprehensive study of rescattering effects in three-body $B$ decays is beyond the scope of the present work.
At any rate, in this work we shall use the phenomenological phase $\delta\approx \pm \pi$ to describe the decays and \CP violation of $B^-\to K^+K^-\pi^-, K^-\pi^+\pi^-$.

\subsubsection{\CP violation in $B^-\to \rho^0\pi^-$}

It has been claimed that the observed large localized \CP violation in $B^-\to \pi^+\pi^-\pi^-$ may result from the
interference of a light scalar meson $f_0(500)$  and the vector $\rho^0(770)$ resonance \cite{Zhang,Bhattacharya}, even
though the latter one is not covered in the low mass region $m^2_{\pi^+\pi^- \rm ~low}<0.4$ GeV$^2$.
Let us consider the intermediate state $\rho^0$ in the $B^-\to\pi^+\pi^-\pi^-$ decay. As shown in Table \ref{tab:Kpipi}, the
calculated
$\B(B^-\to \rho^0\pi^-)=(7.3\pm0.4)\times 10^{-6}$ is consistent with the world average $(8.3^{+1.2}_{-1.3})\times
10^{-6}$ \cite{HFAG} within errors.
Its \CP asymmetry is found to be $\A_{C\!P}(\rho^0\pi^-)=0.059^{+0.012}_{-0.010}$. At first sight, this seems to be in
agreement in sign with the BaBar measurement $0.18\pm0.07^{+0.05}_{-0.15}$ from the Dalitz plot analysis of $B^-\to \pi^+\pi^-\pi^-$ \cite{BaBarpipipi}. However, theoretical
predictions based on QCDF, pQCD and SCET all lead to a negative \CP asymmetry of order $-0.20$ for $B^-\to
\rho^0\pi^-$ (see Table XIII of \cite{CC:Bud}). As shown explicitly in Table IV of \cite{CC:Bud}, within the
framework of QCDF, the inclusion of $1/m_b$ power corrections to penguin annihilation is responsible for the sign flip of
$\A_{C\!P}(\rho^0\pi^-)$ to a negative one.  Specifically, we shall use
\be \beta_3^p[\pi\rho]=-0.03+0.02i, \qquad \beta_3^p[\rho\pi]=0.004-0.049i,
\en
for $p=u,c$.
While the tree-dominated $B^-\to \rho^0\pi^-$ rate is affected only slightly by the power correction, \CP asymmetry flips the sign and becomes $-0.21$. From Table \ref{tab:3pi} we see that the inclusive and regional \CP asymmetries induced by resonances now become negative. Consequently, the predicted $\A_{C\!P}^{\rm incl}$ is wrong in sign, while $\A_{C\!P}^{\rm low}$ and $\A_{C\!P}^{\rm resc}$ are too small when compared with experiment. Hence, the LHCb data imply positive \CP violation induced by the $\rho$ and $f_0$ resonances. Indeed, LHCb has measured asymmetries in $B^-\to\pi^+\pi^-\pi^-$ in four distinct regions dominated by the $\rho$ \cite{LHCb:2014}: I: $0.47<m(\pi^+\pi^-)_{\rm low}<0.77$ GeV, $\cos\theta>0$, II: $0.77<m(\pi^+\pi^-)_{\rm low}<0.92$ GeV, $\cos\theta>0$,
III: $0.47<m(\pi^+\pi^-)_{\rm low}<0.77$ GeV, $\cos\theta<0$, and IV: $0.77<m(\pi^+\pi^-)_{\rm low}<0.92$ GeV, $\cos\theta<0$. It is seen that $\A_{C\!P}$ changes sign at $m(\pi^+\pi^-)\sim m_\rho$. Summing over the regions I-IV yields \CP asymmetry consistent with zero with slightly positive central value (see Table IV of \cite{LHCb:2014}).

%%%%%%%%%%%%%%%%%%%%%%%%%%%%%%%%%%%
\begin{table}[t]
\caption{Predicted inclusive and regional \CP asymmetries (in \%) in $B^-\to\pi^+\pi^-\pi^-$ decay when penguin annihilation is added to render  $\A_{C\!P}(\rho^0\pi^-)\approx -0.21$.  }
\begin{ruledtabular} \label{tab:3pi}
\begin{tabular}{l  c c c c }
  & NR & RES & NR+RES & Expt
  \\ \hline
$\A_{C\!P}^{\rm incl}$
  & $25.0^{+4.4+2.1+0.0}_{-2.7-3.1-0.1}$
  & $-16.3^{+0.0+1.5+0.0}_{-0.0-1.0-0.0}$
  & $-6.7^{+1.6+1.5+0.0}_{-2.6-1.3-0.0}$
  & $5.8\pm2.4$ \\
$\A_{C\!P}^{\rm low}$
  & $58.3^{+3.6+2.6+0.8}_{-3.7-4.0-0.8}$
  & $-16.8^{+0.0+1.5+0.0}_{-0.0-1.1-0.0}$
  & $6.0^{+0.2+3.1+0.0}_{-0.4-1.2-0.0}$
  & $58.4\pm9.7$ \\
$\A_{C\!P}^{\rm resc}$
  & $36.7^{+6.2+3.2+0.1}_{-3.7-4.6-0.2}$
  & $-11.4^{+0.0+1.5+0.0}_{-0.0-1.0-0.0}$
  & $0.4^{+1.2+2.0+0.0}_{-2.1-1.8-0.0}$
  & $17.2\pm2.7$ \\
\end{tabular}
\end{ruledtabular}
\end{table}
%%%%%%%%%%%%%%%%%%%%%%%%%%%%%%%%

Therefore, we encounter a puzzle here. On one hand, BaBar and LHCb measurements of $B^-\to \pi^+\pi^-\pi^-$ seem to indicate a 
positive \CP asymmetry in the $m(\pi^+\pi^-)$ region peaked at $m_\rho$. On the other hand, all theories predict a large and negative \CP violation in $B^-\to \rho^0\pi^-$. This issue concerning $\A_{C\!P}(\rho^0\pi^-)$ needs to be resolved.

\subsubsection{Local \CP violation in other invariant mass regions}

\begin{figure}[t]
\centering
    \includegraphics[scale=0.6]{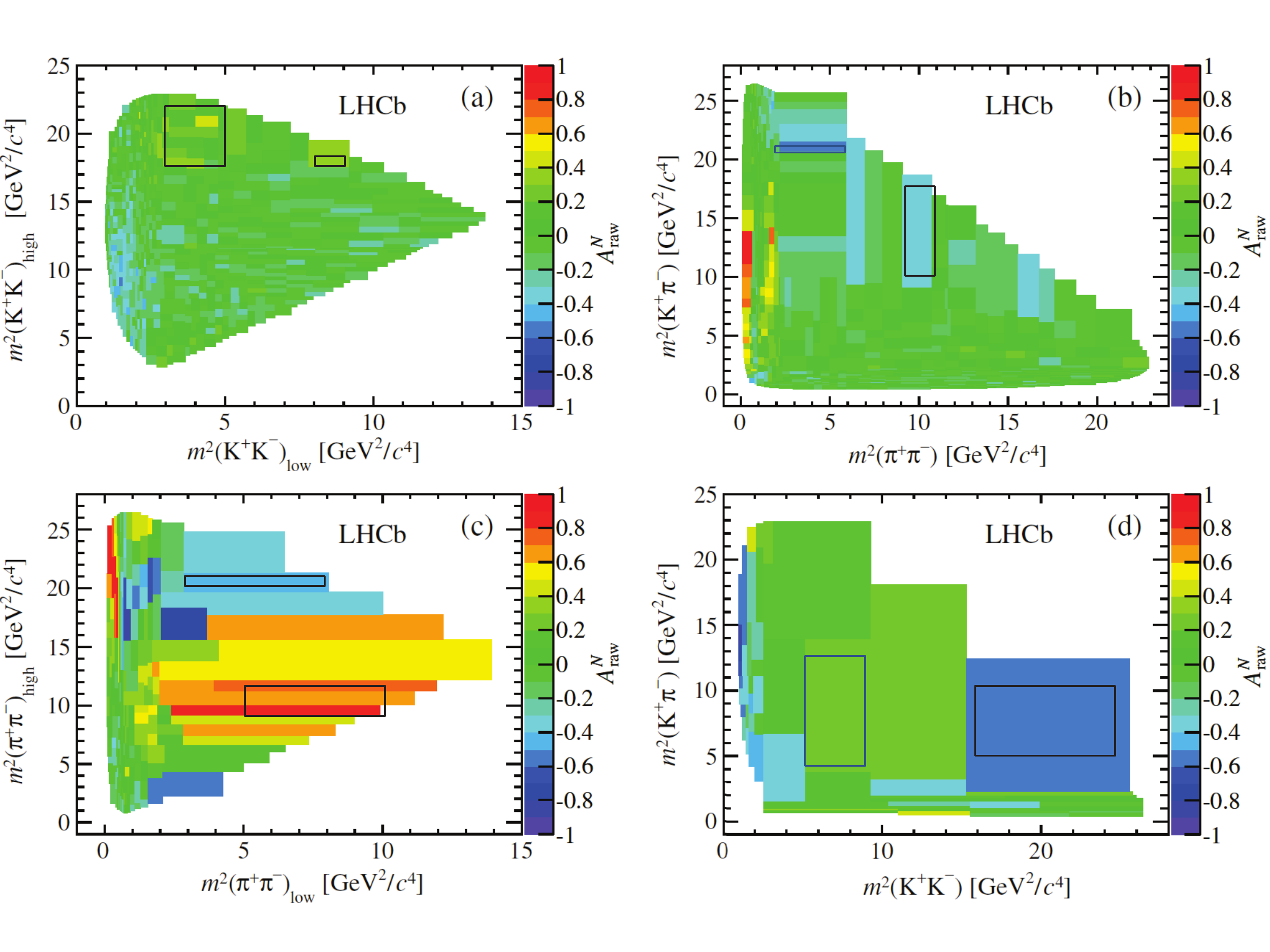}
\caption{ Local \CP asymmetry distributions in the invariant mass regions depicted by the black rectangles for (a) $B^\pm\to K^\pm K^+K^-$,  (b) $B^\pm\to K^\pm \pi^+\pi^-$, (c) $B^\pm\to \pi^\pm\pi^+\pi^-$, and (d) $B^\pm\to \pi^\pm K^+ K^-$. Dalitz plots of \CP-asymmetry distributions are taken from \cite{LHCb:2014}. }
\label{fig:localCP}
\end{figure}

For regional \CP violation, so far we have focused on the small invariant mass region specified in Eq. (\ref{eq:KKKlocalCP}) and the rescattering region of $m_{\pi\pi}$ and $m_{K\!\bar K}$ between 1.0 and 1.5 GeV. As noticed in passing,
the magnitude and sign of \CP asymmetries in the Dalitz plot vary from region to region. A successful model must explain not only the inclusive asymmetry but also regional \CP violation. Therefore, the measured {\it CP}-asymmetry Dalitz distributions put stringent constraints on the models.
In the following we consider the distribution of $\A_{C\!P}$ in some (large) invariant mass regions to test our model.

\vskip 0.2cm
\noindent \underline{$B^\pm\to K^\pm K^+K^-$}

We see from Fig. \ref{fig:localCP}(a) that $\A_{C\!P}$ is mostly negative in the Dalitz plot region with $m(K^+K^-)_{\rm low}$ between 1 and 1.6 GeV and $m(K^+K^-)_{\rm high}$ below 4 GeV, but it can be positive at $m(K^+K^-)_{\rm high}>4$ GeV (see also Fig. 2 of \cite{LHCb:2016}). We consider two regions with positive $\A_{C\!P}$: (i) $m^2(K^+K^-)_{\rm low}=3$--$5~{\rm GeV}^2$ and $m^2(K^+K^-)_{\rm high}=18$--$22~{\rm GeV}^2$, and (ii) $m^2(K^+K^-)_{\rm low}=8$--$9~{\rm GeV}^2$ and $m^2(K^+K^-)_{\rm high}=18$--$19~{\rm GeV}^2$.
We obtain the values of $\A_{C\!P}$ to be $0.11$ and $0.41$, respectively, in our model. This is consistent with the data as $\A_{C\!P}$ in region (ii) should be much larger than that in region (i).

\vskip 0.2cm
\noindent \underline{$B^\pm\to K^\pm \pi^+\pi^-$}

While the integrated $\A_{C\!P}^{\rm incl}$ is positive in this decay,
Fig. \ref{fig:localCP}(b) shows the distribution of negative \CP asymmetry in the regions such as (i) $m^2(\pi^+\pi^-)=9.5$--$10.5~{\rm GeV}^2$ and $m^2(K^+\pi^-)=10$--$18~{\rm GeV}^2$  and (ii) $m^2(\pi^+\pi^-)=2$--$6~{\rm GeV}^2$ and $m^2(K^+\pi^-)=20.5$--$21.5~{\rm GeV}^2$. Our model leads to $\A_{C\!P}^{\rm local}\approx -0.09$ and $-0.04$, respectively. Experimentally,
$|\A_{C\!P}|$ in region (ii) should be larger. Therefore, while the sign is correctly predicted, the relative magnitude of $\A_{C\!P}$ in regions (i) and (ii) is not borne out by experiment.

\vskip 0.2cm
\noindent \underline{$B^\pm\to\pi^\pm\pi^+\pi^-$}

It is obvious from Fig. \ref{fig:localCP}(c) that $\A_{C\!P}$ is very large and positive in the region of $5< m^2(\pi^+\pi^-)_{\rm low}<10~{\rm GeV}^2$ and $9< m^2(\pi^+\pi^-)_{\rm high}<12~{\rm GeV}^2$, and it becomes negative in the region of  $3<m^2(\pi^+\pi^-)_{\rm low}<8~{\rm GeV}^2$ and $20< m^2(\pi^+\pi^-)_{\rm high}<21~{\rm GeV}^2$. We obtain $\A_{C\!P}^{\rm local}\approx 0.47$ and $-0.29$, respectively, in qualitative agreement with experiment.

\vskip 0.2cm
\noindent \underline{$B^\pm\to \pi^\pm K^+K^-$}

Fig. \ref{fig:localCP}(d) shows that $\A_{C\!P}$ is large and negative in the region of (i) $16<m^2(K^+K^-)<25~{\rm GeV}^2$ and $5<m^2(K^+\pi^-)<10~{\rm GeV}^2$. It changes sign in the region of (ii) $5< m^2(K^+K^-)<9~{\rm GeV}^2$ and $4< m^2(K^+\pi^-)<13~{\rm GeV}^2$. Our results $\A_{C\!P}^{\rm local}\approx 0.36$ and $-0.44$ in regions (i) and (ii), respectively, are not consistent with experiment. If the phase $\delta$ is set to zero, we will have $\A_{C\!P}^{\rm local}\approx -0.73$ and $0.54$, respectively, in qualitative agreement with the data. Thus it is possible that the phase $\delta$ is energy dependent and it vanishes in the large invariant mass region. This issue is currently under study.

\vskip 0.2 cm
In short, for local \CP asymmetries in  various (large) invariant mass regions, our model predictions are in qualitative agreement with experiment for $K^+K^-K^-$ and $\pi^+\pi^-\pi^-$ modes and yield a correct sign for $K^-\pi^+\pi^-$. However,
it appears that the phase $\delta$ needs to vanish in the large invariant mass region for $K^+K^-\pi^-$ in order to accommodate the observation.

\section{Comparison with other works}

\CP violation in three-body decays of the charged $B$ meson has been investigated in
Ref. \cite{Zhang,Zhang:2013iga,Bhattacharya,XGHe,Cheng:2013dua,Bediaga,Lesniak,Li:2014,Wang:2014ira,Bhattacharya:2014,Krankl:2015fha,Wang:2015ula,Bediaga:2015,Bediaga:2015mia}.
The authors of \cite{Zhang,Bhattacharya}  considered the possibility of having a large local \CP violation in $B^-\to \pi^+\pi^-\pi^-$ resulting from the interference of the resonances $f_0(500)$ and  $\rho^0(770)$. A similar mechanism has been applied to the decay $B^-\to K^-\pi^+\pi^-$ \cite{Zhang:2013iga}.
Studies of flavor SU(3) symmetry imposed on the decay amplitudes and its implication on \CP violation were elaborated on in \cite{XGHe,Bhattacharya:2014}.
The observed \CP asymmetry in $B^-\to\pi^+\pi^-\pi^-$ decays changes sign at a value of $m(\pi^+\pi^-)_{\rm low}$ close to the $\rho(770)$ resonance \cite{LHCb:2014}. It was argued in \cite{Wang:2014ira} that the sign change is caused by the $\rho$--$\omega$ mixing.
In our work, we have taken into account both resonant and nonresonant amplitudes simultaneously and worked out their contributions to branching fractions and \CP violation in details. We found that even in the absence of $f_0(500)$ resonance, local \CP asymmetry in $\pi^+\pi^-\pi^-$ can already reach the level of 17\% due to nonresonant and other resonant contributions. Moreover, the regional asymmetry induced solely by the nonresonant component can be as large as 58\% in our calculation. In our work and also in the work of \cite{Bediaga,Bediaga:2015} to be discussed below, the sign change is ascribed to the real part of the Breit-Wigner propagator for the $\rho(770)$ resonance.

Based on the constraint of ${\it CP\!T}$ invariance on final-state interactions, the authors of \cite{Bediaga,Bediaga:2015} have studied \CP violation in charmless three-body charged $B$ decays, especially the \CPP-asymmetry distribution in the mass region below 1.6 GeV. We first recapitulate the main points of this work.
Writing the $S$ matrix as $S_{\lambda'\lambda}=\delta_{\lambda'\lambda}+it_{\lambda'\lambda}$ and the decay amplitude to the leading order in $t$ as
\be \label{eq:Ares}
{\cal A}(h\to\lambda)=A_\lambda+e^{-i\gamma}B_\lambda+i\sum_{\lambda'}t_{\lambda'\lambda}
(A_{\lambda'}+e^{-i\gamma}B_{\lambda'}),
\en
with $A_\lambda$ and $B_\lambda$ being complex amplitudes invariant under {\it CP},
it follows that the rate difference reads \cite{Bediaga,Bediaga:2015}
\be
\Delta\Gamma_\lambda &\equiv& \Gamma(h\to\lambda)-\Gamma(\bar h\to \bar\lambda) \non \\
&=& 4(\sin\gamma){\rm Im}(B^*_\lambda A_\lambda)+ 4(\sin\gamma)\sum_{\lambda'}{\rm Re}[B^*_\lambda t_{\lambda'\lambda}A_{\lambda'}-
B^*_{\lambda'} t_{\lambda'\lambda}^*A_{\lambda}] \non\\
&= & \Delta\Gamma_\lambda^{\rm SD}+\Delta\Gamma_\lambda^{\rm FSI},
\en
where the first term corresponds to the familiar short-distance contribution to direct \CP asymmetry and the second term arises from final-state rescattering (so-called compound \CP violation).
It is interesting to notice the relation (see \cite{Bediaga:2015} for the derivation)
\be \label{eq:CPT}
\sum_\lambda\Delta \Gamma_\lambda^{\rm FSI}=0
\en
is valid irrespective of the short-distance one. When the ${\it CP\!T}$ condition  $\sum_\lambda{\rm Im}[B^*_\lambda A_\lambda]=0$  is imposed, the ${\it CP\!T}$ constraint $\sum_\lambda \Delta\Gamma_\lambda=0$ follows.

Suppose only the two channels $\alpha=\pi^+\pi^-P^-$ and $\beta=K^+K^-P^-$ ($P=\pi,K$) in $B^-$ decays are strongly coupled through strong interactions with the third meson $P$ being treated as a bachelor or a spectator, it follows from Eq. (\ref{eq:CPT}) that $\Delta\Gamma^{\rm FSI}_\alpha=-\Delta\Gamma^{\rm FSI}_\beta$ (not $\Delta\Gamma_\alpha=-\Delta\Gamma_\beta$!). It should be stressed again that this relation is not imposed by hand, rather it is a consequence of the assumption of only two channels coupled through final-state resacttering. As a result,
\be \label{eq:ACPratio}
\left( \A_{C\!P}^{\rm incl}(K^+K^-\pi^-)\over \A_{ C\!P}^{\rm incl}(\pi^+\pi^-\pi^-) \right)^{\rm FSI} &=& -{\B(B^-\to \pi^+\pi^-\pi^-)\over \B(B^-\to K^+K^-\pi^-)}= -3.0\pm 0.5,  \non \\
\left( \A_{C\!P}^{\rm incl}(K^+K^-K^-)\over \A_{C\!P}^{\rm incl}(\pi^+\pi^-K^-) \right)^{\rm FSI} &=& -{\B(B^-\to K^+K^-K^-)\over \B(B^-\to \pi^+\pi^-K^-)}= -1.5\pm 0.1,
\en
where we have used the branching fractions listed in Table \ref{tab:Kpipi} and the averaged ones: $\B(B^-\to \pi^+\pi^-K^-)=(51.0\pm2.9)\times 10^{-6}$ and $\B(B^-\to K^+K^-K^-)=(33.0\pm1.0)\times 10^{-6}$. Experimentally, the ratios in Eq. (\ref{eq:ACPratio}) are measured to be of order $-2.1$ and $-1.4$, respectively. The coincidence between theory and experiment suggests that the LHCb data of \CP asymmetries could be described in terms of final-state rescattering.
For three-body $B$ decays,
the strong couplings between $K^+K^-$ and $\pi^+\pi^-$ channels with the ${\it CP\!T}$ constraint were used in \cite{Bediaga:2015}  to fit the observed
asymmetries in some channels and then predict \CP violation in other modes. Explicitly, the amplitude Eq. (\ref{eq:Ares}) is fitted to the LHCb data of the distribution of \CP asymmetries in $m(\pi^+\pi^-)$ measured in $B^-\to \pi^+\pi^-P^-$ decays with $P=\pi,K$. Then the fit parameters in $\Delta\Gamma_\alpha^{\rm FSI}$ are used to predict the $\Delta\Gamma^{\rm FSI}_\beta(s)$ distributions of $B^-\to K^+K^-P^-$ decays in $m(K^+K^-)$ (see Figs. 10 and 12 of \cite{Bediaga:2015}). It turns out that the \CPP-asymmetry distributions of $B^-\to K^+K^-P^-$ observed by LHCb in the rescattering region are fairly accounted for by the final-state rescattering of $\pi^+\pi^-\leftrightarrow K^+K^-$.

In short, final-state interactions play an essential role in the work of \cite{Bediaga,Bediaga:2015}. The ${\it CP\!T}$ relation  $\Delta\Gamma^{\rm FSI}_\alpha=-\Delta\Gamma^{\rm FSI}_\beta$ is used to describe \CPP-asymmetry distributions in $B^-\to K^+K^-P^-$ decays after a fit to $B^-\to \pi^+\pi^-P^-$ channels. Final-state rescattering of $\pi^+\pi^-\leftrightarrow K^+K^-$ dominates the asymmetry in the mass region between 1 and 1.5 GeV.
On the contrary, we performed a dynamical model calculation of partial rates and \CP asymmetries without taking into account final-state interactions explicitly. We accentuate the crucial role played by nonresonant contributions. Our predicted inclusive \CP asymmetries
for $\pi^+\pi^-\pi^-$ and $K^+K^-K^-$ agree with experiment and have nothing to do with $\pi^+\pi^-$ and $K^+K^-$ final-state rescattering, while the calculated \CP asymmetries for $K^+K^-\pi^-$ and $\pi^+\pi^-K^-$ are wrong in sign. Hence, we introduce an additional strong phase $\delta$ to flip the sign.

\section{Conclusions}
We have presented in this work a study of charmless three-body decays of $B$ mesons using a simple
model based on the factorization approach. Our main results are:

\begin{itemize}

\item Dominant nonresonant contributions to tree-dominated and penguin-dominated three-body decays arise
from the $b\to u$ tree transition and $b\to s$ penguin transition, respectively. The former can be evaluated in the framework of  heavy meson chiral perturbation theory supplemented by some energy dependence to ensure that HMChPT results are valid in chiral limit. The latter is
governed by the matrix element of the scalar density $\la M_1M_2|\bar q_1 q_2|0\ra$.

\item Based on the factorization approach, we have considered the
resonant contributions to three-body decays and computed the rates
for the quasi-two-body decays $B\to VP$ and $B\to SP$. While the
calculated branching fractions for the tree-dominated modes such as $\rho\pi$ and $f_0(980)\pi$ are
consistent with experiment, the predicted rates for penguin-dominated  $\phi
K,~K^*\pi,~\rho K$ and $K_0^*(1430)\pi$ channels are too small compared to
the data. This implies the importance of power corrections. We follow the QCD factorization approach to introduce the penguin annihilation characterized by the parameter $\beta_3$ to improve the discrepancy between theory and experiment for penguin-dominated ones.

\item  The branching fraction of nonresonant contributions is of order $(15-20)\times 10^{-6}$ in penguin-dominated
decays $B^-\to K^+K^-K^-,K^-\pi^+\pi^-$ and of order $(3-5)\times 10^{-6}$ in tree-dominated decays $B^-\to \pi^+\pi^-
\pi^-, K^+K^-\pi^-$. The nonresonant fraction is predicted to be around 55\% for the $B^-\to K^+K^-\pi^-$ decay.

\item We have updated the predictions for the resonant and nonresonant contributions to  $B^-\to \ov K^0\pi^-\pi^0$, $B^-\to K^-\pi^0\pi^0$, $\ov B^0\to\ov K^0\pi^+\pi^-$ and $\ov B^0\to K^-\pi^+\pi^0$. The calculated total branching fractions are smaller than experiment. This is ascribed to the fact that the predicted $B\to K_0^*(1430)\pi$ rates in factorization or QCDF are too small compared to the data and that the $K_0^*(1430)$ has the largest contributions to $B\to K\pi\pi$ decays.

\item
  In our study of $B^-\to \pi^-\pi^+\pi^-$, we find that $\A_{C\!P}(\rho^0\pi^-)$ is positive. Indeed, both BaBar and LHCb measurements of $B^-\to \pi^+\pi^-\pi^-$ indicate positive \CP asymmetry in the $m(\pi^+\pi^-)$ region peaked at $m_\rho$. On the other hand, all theories predict a large and negative \CP violation in $B^-\to \rho^0\pi^-$. We have shown that if we add $1/m_b$ penguin-annihilation induced power correction to render $\A_{C\!P}(\rho^0\pi^-)$ negative, $\A_{C\!P}^{\rm incl}$ will be wrong in sign and the predicted regional \CP asymmetries will become too small compared to experiment.
  Therefore, the issue with \CP violation in $B^-\to\rho^0\pi^-$  needs to be resolved.

\item While the calculated direct \CP asymmetries for $K^+K^-K^-$ and $\pi^+\pi^-\pi^-$ modes are in good agreement
with experiment in both magnitude and sign, the predicted asymmetries in $B^-\to \pi^- K^+K^-$ and $B^-\to K^-\pi^
+\pi^-$ are wrong in signs when confronted with experiment. This is attributed to the sizable nonresonant contributions which are opposite in sign to the experimental measurements (see Table \ref{tab:CPdelta}). We have studied final-state inelastic $\pi^+\pi^-\leftrightarrow K^+K^-$ rescattering and found that \CP violation for both $\pi^+\pi^-\pi^-$ and $K^+K^-K^-$ is heading to the wrong direction, making the discrepancy even worse. In order to accommodate the branching fraction of nonresonant component and \CP asymmetry observed in $B^-\to K^-\pi^+\pi^-$, the matrix element $\la K\pi|\bar sq|0\ra$ should have an extra strong phase $\delta$ of order $\pm\pi$ in addition to the phase characterized by the parameter $\sigma_{\rm NR}$. This phase $\delta$ may arise from some sort of power corrections such as final-state interactions. The matrix element $\la K\pi|\bar qs|0\ra$ relevant to the decay $B^-\to \pi^-K^+K^-$ is related to $\la K\pi|\bar sq|0\ra$ via $U$-spin symmetry.

\item In this work, there are three sources of strong phases: effective Wilson coefficients, propagators of resonances and the
matrix element of scalar density $\la M_1M_2|\bar q_1q_2|0\ra$.  There are two sources for the phase in the penguin matrix element of scalar densities: $\sigma_{\rm NR}$ and $\delta$ for $K\pi$--vacuum matrix elements.

\item
  Nonresonant \CP violation is usually much larger than the resonant one and the interference effect between resonant and nonresonant components is generally quite significant. If nonresonant contributions are turned off in the $B^-\to K^+K^-K^-$ mode, the predicted \CP asymmetries due to resonances will be incorrect in sign. Since this decay is predominated by the nonresonant background, the magnitude and the sign of its \CP asymmetry should be governed by the nonresonant term.

\item
  We have studied {\it CP}-asymmetry Dalitz distributions in some (large) invariant mass regions to test our model. Our model predictions are in qualitative agreement with experiment for $K^+K^-K^-$ and $\pi^+\pi^-\pi^-$ modes and yield a correct sign for $K^-\pi^+\pi^-$. However,
  it appears that the phase $\delta$ needs to vanish in the large invariant mass region for $K^+K^-\pi^-$ in order to accommodate the observation.

\end{itemize}

\vskip 2.0cm \acknowledgments
This work was supported in part by the Ministry of Science and Technology of Taiwan under Grant Nos.~MOST~104-2112-M-001-022 and 103-2112-M-033-002-MY3 and by the National Natural Science Foundation of China under Grant No. 11347030, the Program of Science and Technology Innovation Talents in Universities of Henan Province 14HASTIT037.

%\newpage

%%%%%%%%%%%%%%%%%%%%%%%%%%%%%%%%%%%%%%%%%%%%%%%%%%%%%%%%

\end{document}